\documentclass[11pt,letterpaper]{article}
\usepackage[margin=1in]{geometry}
\usepackage[T1]{fontenc}
\usepackage{lmodern,microtype,needspace}
\usepackage{amsmath,amssymb,amsthm,mathtools}
\usepackage{graphicx,float}
\usepackage{enumitem,booktabs,array,tikz}
\usetikzlibrary{arrows.meta,patterns}
\usepackage{algorithm2e}
\usepackage[colorlinks=true,linkcolor=blue!55!black,citecolor=blue!55!black,
            urlcolor=blue!55!black,bookmarksnumbered=true]{hyperref}
\usepackage[capitalise,nameinlink]{cleveref}

\newtheorem{theorem}{Theorem}
\newtheorem{lemma}{Lemma}

\newtheorem{corollary}{Corollary}
\theoremstyle{definition}
\newtheorem{definition}{Definition}
\newtheorem{observation}{Observation}
\newtheorem{claim}{Claim}
\theoremstyle{remark}

\crefname{equation}{}{}
\crefname{appendix}{Appendix}{Appendices}
\crefname{claim}{Claim}{Claims}
\crefname{observation}{Observation}{Observations}

\newcommand{\E}{\mathbb E}
\newcommand{\R}{\mathbb R}
\newcommand{\Q}{\mathbb Q}

\newcommand{\one}{\mathbf 1}
\newcommand{\Opt}[1]{\text{\normalfont\scshape Opt}(#1)}
\newcommand{\Optf}[1]{\text{\normalfont\scshape Opt}_{\mathrm{frac}}(#1)}
\newcommand{\Mech}[2]{\text{\normalfont\scshape Mech}_{#1}(#2)}

\hypersetup{pdftitle={Tight bounds for fractional and randomized truthful scheduling},
 pdfauthor={Georgios Anastasiadis; George Christodoulou; Elias Koutsoupias; Annamaria Kovacs; Conrad Schecker}}

\title{Tight bounds for fractional and randomized\\[0.3em]truthful scheduling}
\author{
 Georgios Anastasiadis\thanks{School of Informatics, Aristotle University of Thessaloniki, Greece. Email: \texttt{ganastag@csd.auth.gr}}\and
 George Christodoulou\thanks{School of Informatics, Aristotle University of Thessaloniki, Greece; Archimedes, Athena Research Center, Greece. Email: \texttt{gichristo@csd.auth.gr}}\and
 Elias Koutsoupias\thanks{Department of Computer Science, University of Oxford, UK. Email: \texttt{elias.koutsoupias@cs.ox.ac.uk}}\and
 Annamária Kovács\thanks{Institute of Computer Science, Goethe University Frankfurt, Germany. Email: \texttt{panni@cs.uni-frankfurt.de}}\and
 Conrad Schecker\thanks{Institute of Computer Science, Goethe University Frankfurt, Germany. Email: \texttt{schecker@em.uni-frankfurt.de}}
}
\date{}

\begin{document}
\hypersetup{pageanchor=false}
\begin{titlepage}
\maketitle
\begin{abstract}

We resolve one of the longest-standing open problems in 
algorithmic mechanism
design by showing that randomization cannot yield a sublinear
approximation for truthful scheduling on unrelated machines,
even when truthfulness is required only in expectation.
Our main technical result concerns fractional scheduling, where
tasks may be split among machines. We prove that every deterministic
truthful fractional mechanism on $n$ unrelated machines has
worst-case approximation ratio at least $(n+1)/2$, even against
the optimal integral makespan. This matches the known upper bound
exactly and settles the optimal approximation ratio of truthful
fractional scheduling.
Since the expected allocations and payments of a
truthful-in-expectation mechanism define a truthful fractional
mechanism, the same lower bound applies to all
truthful-in-expectation mechanisms and, in particular, to all
universally truthful mechanisms.
On the positive side, we give a universally truthful mechanism
with approximation ratio $n/2+o(n)$. Together, these results
establish the optimal leading constant $1/2$ for randomized
truthful scheduling. Thus, randomization improves on the optimal
deterministic approximation ratio by an asymptotic factor of two,
and no more.

\end{abstract}
\thispagestyle{empty}
\end{titlepage}
\setcounter{page}{1}
\hypersetup{pageanchor=true}

\section{Introduction}

We study truthful fractional and randomized mechanisms for the classic
problem of scheduling on unrelated machines. In this problem, there are
$m$ tasks to be assigned to $n$ machines, and each machine $i$ has a
processing time $t_{ij}$ for executing task $j$. The objective is to
minimize the \emph{makespan}, defined as the maximum completion time
among all machines.

Unlike traditional algorithms, a truthful mechanism must incentivize
each machine $i$ by appropriate payments to reveal its \emph{private
  values} $t_{ij}$. Nisan and Ronen, in their seminal
work~\cite{NisanRonen2001}, laid the foundations of
\emph{algorithmic mechanism design}. They proposed the scheduling
problem as the canonical algorithmic problem in this area and posed
the question of the best approximation ratio that can be achieved by
truthful deterministic and randomized mechanisms. Two aspects make
this a challenging question: each machine's private input is
\emph{multidimensional}, and the objective is of \emph{min-max} type,
unlike the min-sum objective truthfully optimized by the VCG mechanism.

For \emph{deterministic} mechanisms, the question was recently settled
by Christodoulou, Koutsoupias and Kov\'acs~\cite{CKK26}, who resolved
the Nisan--Ronen conjecture: the best approximation ratio of truthful
mechanisms is exactly $n$, attained by VCG.

In light of this negative result, relaxing the model is the natural
next step, and \emph{two relaxations} have been studied: \emph{randomized and
fractional mechanisms}. Both provably help: for two machines, where no
deterministic truthful mechanism achieves a ratio below~$2$, Nisan and
Ronen~\cite{NisanRonen2001} gave a universally truthful mechanism with
ratio $7/4$, and the fractional mechanism of Christodoulou, Koutsoupias
and Kov\'acs~\cite{CKK10} achieves $3/2$.
Randomized mechanisms come in two flavors: universally truthful and
truthful-in-expectation. A randomized
mechanism is called \textit{universally truthful} if it is defined as
a probability distribution over deterministic truthful mechanisms. A
randomized mechanism is \emph{truthful-in-expectation} when agents
cannot strictly improve their expected utility by misreporting. The class of
truthful-in-expectation mechanisms is broader, but the universally
truthful mechanisms have the important characteristic that they are
truthful even ex post (that is, even when the random coins become
known). In the model of \emph{fractional} mechanisms~\cite{CKK10}, each
task can be fractionally allocated to multiple machines, and each
machine's processing cost is linear in its assigned fractions. When it
is measured against the integral optimum, as randomized mechanisms are,
this is an even stronger relaxation: the expected allocations and
payments of a truthful-in-expectation mechanism form a truthful
fractional mechanism, whose makespan is at most the expected makespan.
Hence a lower bound for fractional mechanisms against the integral
optimum also holds for truthful-in-expectation mechanisms, and in
particular for universally truthful ones.

For fractional mechanisms, Christodoulou, Koutsoupias and
Kov\'acs~\cite{CKK10} gave a truthful mechanism with approximation ratio
$(n+1)/2$, and showed, as did Lu and Yu~\cite{LY08}, that no
\emph{task-independent} mechanism does better. For general mechanisms,
however, the best known lower bound was $2-1/n$, for both fractional and
randomized mechanisms~\cite{MualemSchapira2018,CKK10}, while the best
known randomized mechanisms achieve $(n+5)/2$ in
expectation~\cite{LY08} and about $0.837n$
universally~\cite{LuYu2008Improved}.

Our main result resolves the question for randomized and fractional mechanisms.

\begin{samepage}
\begin{theorem}\label{thm:nrtheorem}
Every deterministic truthful fractional mechanism for scheduling $n$
unrelated machines has approximation ratio at least $(n+1)/2$, even
against the optimal \emph{integral} makespan, when arbitrary finite
numbers of tasks are allowed.
\end{theorem}
\end{samepage}

Together with the mechanism of~\cite{CKK10}, this shows that the optimal
approximation ratio of truthful fractional scheduling is exactly
$(n+1)/2$. 

A truthful-in-expectation mechanism is measured against the \emph{integral
optimum}, which can exceed the fractional one. But since the benchmark in
\cref{thm:nrtheorem} is the integral optimum, the bound applies to
randomized mechanisms as well.

\begin{corollary}\label{cor:intro-tie}
Every truthful-in-expectation mechanism for scheduling $n$ unrelated
machines, with integral or fractional allocations, has expected
approximation ratio at least $(n+1)/2$. In particular, this holds for
every universally truthful mechanism.
\end{corollary}

This is the first superconstant lower bound for general randomized
truthful mechanisms: previously, the best lower bound was
$2-1/n$~\cite{MualemSchapira2018}.

On the positive side, we give a universally truthful mechanism,
\textnormal{\textsc{Exp-Bounded-Square}}, with approximation ratio
$(n+1)/2+O(\sqrt{n\log n})=n/2+o(n)$. It adapts the
\textnormal{\textsc{Bounded-Square}} mechanism of Lu and Yu~\cite{LY08}
and improves the previous best linear coefficient of about
$0.837$~\cite{LuYu2008Improved}. 

\paragraph{Conclusion.} Since Nisan and Ronen posed the question of randomized mechanisms for
scheduling~\cite{NisanRonen2001}, the main hope has been that
randomization could overcome the limits of deterministic mechanisms: perhaps randomized mechanisms could achieve a
constant, or at least sublinear, approximation ratio. For more than two
decades this hope remained open, as the best lower bound for randomized
mechanisms stayed constant. Taken together, our results settle the
question. The lower bound becomes linear, $(n+1)/2$, and it is tight:
exactly for fractional mechanisms, up to an additive~$2$ for
truthful-in-expectation mechanisms~\cite{LY08}, and up to an additive
$O(\sqrt{n\log n})$ for universally truthful ones. Randomization thus improves on the
deterministic optimum~$n$ asymptotically by a factor of two, and no more.

\paragraph{Techniques.}
The bound $(n+1)/2$ was known for task-independent
mechanisms~\cite{CKK10,LY08}; the difficulty is that a general mechanism
can couple the allocations of different tasks arbitrarily.

We use graph instances: the machines are the vertices, and each task is
either an edge between the only two machines that can process it at
reasonable cost, or a self-loop that only its owner can process.
Edges between the same two machines are called parallel, and a
star consists of one edge from a root to each other machine.
Our proof uses many parallel edges between each pair of machines.
Averaging over task permutations within each pair, and over the numbers
of added \emph{reservoir} tasks with a zero-cost machine, preserves truthfulness and approximation. In a
suitable limit, a task with a unique zero-cost machine goes wholly to
that machine, and changing a task between types with a zero-cost machine
does not affect the allocation of the other tasks. Truthfulness of
\emph{both} endpoints then makes the disutility functions additive
across parallel edges. An exact identity for these additive and symmetric
disutility functions shows that, for a suitable root and a fixed star,
the root retains almost half of the star's tasks. With a unit-cost
self-loop at the root, the root's load approaches $(n+1)/2$, while the
optimum approaches one. \Cref{sec:overview} gives a more detailed
overview.\footnote{Generative AI assisted the development and exposition of the lower-bound proof. Its role is described in the AI methodology statement at the end of the paper.}

The remainder of the paper is organized as follows. \Cref{sec:prelim}
formally defines the scheduling problem, the mechanism design framework,
and disutility functions. \Cref{sec:lb} proves the lower bound for
fractional mechanisms and derives its consequences for randomized
mechanisms. \Cref{sec:upperbound} presents the
\textnormal{\textsc{Exp-Bounded-Square}}
mechanism and proves its universal truthfulness and approximation
guarantee. Appendices~\ref{app:padding}--\ref{app:laplace}
contain the proofs of the padding lemma, of two-task separation
without continuity, and of the uniqueness of the Laplace transform.

\subsection{Related Work}

\paragraph{Deterministic Mechanisms.}
The seminal work of Nisan and Ronen~\cite{NisanRonen2001} established a lower bound of $2$ for deterministic truthful mechanisms and showed that the VCG mechanism yields an $n$-approximation, leaving a huge gap. A sequence of works improved the lower bound only incrementally: first to $2.41$~\cite{ChristodoulouKoutsoupiasVidali2009}, then to $2.61$~\cite{KoutsoupiasVidali2013}, later to $2.75$~\cite{GiannakopoulosHammerlPocas2020}, and eventually to $3$~\cite{DobzinskiShaulker2020}. Christodoulou, Koutsoupias, and Kov\'acs~\cite{CKK26} fully closed this gap by showing that no deterministic truthful mechanism can circumvent the $n$-approximation barrier, thereby resolving the long-standing Nisan--Ronen conjecture.

\paragraph{Randomized Mechanisms.}
Mu'alem and Schapira~\cite{MualemSchapira2018} showed a lower bound of $2-1/n$ for both truthful-in-expectation and universally truthful randomized mechanisms. Christodoulou, Koutsoupias, and Kov\'acs~\cite{CKK10} extended this lower bound to fractional mechanisms, where each task can be fractionally allocated to multiple machines. They also presented a (deterministic) fractional mechanism with a guarantee of $(n+1)/2$. Building on this framework, Lu and Yu~\cite{LY08} obtained a truthful-in-expectation mechanism achieving an approximation guarantee of $(n+5)/2$.

Regarding universally truthful mechanisms, Nisan and Ronen~\cite{NisanRonen2001} proposed a mechanism for the case of two machines achieving a $7/4$-approximation. Mu'alem and Schapira~\cite{MualemSchapira2018} extended this mechanism to $n$ machines with an approximation guarantee of $0.875n$, and Lu and Yu~\cite{LuYu2008Improved} subsequently improved its linear coefficient to about $0.837$. The case of two machines has received particular attention~\cite{LY08,ChenDuZuluaga2015}.

\section{Preliminaries}
\label{sec:prelim}

We consider the scheduling problem for unrelated machines. There is a set
$M$ of $m$ tasks that need to be scheduled on a set $N$ of $n$ machines.
Throughout the paper, we write $k=n-1$.
Each task $j\in M$ has a processing time (or value) $t_{ij}\in\R_{\geq0}$
for every machine $i\in N$. The goal is to allocate the tasks so that the
maximum completion time over the machines, called the \emph{makespan}, is
minimized. Each machine $i$ is controlled by a selfish agent, and
the vector $t_i=(t_{i1},\ldots,t_{im})$, the \emph{type} of agent~$i$, is
privately known to that agent. We use the terms machine and agent
interchangeably. We use \emph{value} for a processing time in an input,
and \emph{report} for a type submitted to the mechanism, including a
possible deviation from the agent's true type.

\paragraph{Notation.}
For a set $A$ of coordinates, $\one_A$ is the vector equal to $1$ on $A$
and $0$ elsewhere, and $\one_j=\one_{\{j\}}$ is the $j$-th unit vector;
$\one$ is the all-ones vector, and $0_A$ is the zero vector on the
coordinates in $A$. We write $[k]=\{1,\ldots,k\}$,
$a\cdot b=\sum_ja_jb_j$, and $\R_{\geq0}$ and $\Q_{\geq0}$ for the
nonnegative reals and rationals.

\subsection{Mechanisms}
\label{sec:mechanisms}

A \emph{deterministic fractional mechanism} assigns a fraction
$x_{ij}(t)\in[0,1]$ of each task $j$ to each machine $i$, with
$\sum_i x_{ij}(t)=1$, and pays $P_i(t)$ to machine $i$.
We call $x$ the \emph{allocation function} and $(x,P)$ the mechanism.
Writing $x_i=(x_{ij})_{j\in M}$, the \emph{utility} of machine $i$ is its
payment minus its processing cost, $P_i(t)-t_i\cdot x_i(t)$. The
mechanism is \emph{truthful} if for every agent $i$, true type $t_i$,
report $b_i$ and opposing reports $t_{-i}$,
\begin{equation}\label{eq:tie}
 P_i(t_i,t_{-i})-t_i\cdot x_i(t_i,t_{-i})
 \geq P_i(b_i,t_{-i})-t_i\cdot x_i(b_i,t_{-i}).
\end{equation}
The \emph{completion time} of machine $i$ is
$t_i\cdot x_i(t)=\sum_j t_{ij}x_{ij}(t)$; we also call it the
\emph{load} of $i$. The makespan $\Mech{x}{t}$ is the maximum completion
time. An allocation is \emph{integral} if $x_{ij}(t)\in\{0,1\}$ for all
$i,j$. A deterministic mechanism in the usual sense is a fractional
mechanism whose allocations are all integral; it is truthful if it
satisfies~\eqref{eq:tie}. A mechanism is \emph{task-independent} if the
allocation of each task depends only on the values for that task.

A \emph{randomized mechanism} $R$ outputs, for every input, a random
integral allocation and random payments. Its makespan $\Mech{R}{t}$ is then a random
variable, and its performance is the expected makespan $\E\,\Mech{R}{t}$.
Let $x_{ij}(t)$ be the probability that $R$ assigns task $j$ to machine
$i$ on input $t$, and let $P_i(t)$ be the expected payment to machine
$i$; we assume that expected payments are finite. The mechanism is
\emph{truthful in expectation (TIE)} if no machine can increase its
expected utility by misreporting. Here $x$ and $P$ are ordinary functions
of the reports, and since expected processing cost is linear in the
probabilities, $R$ is TIE exactly when $(x,P)$
satisfies~\eqref{eq:tie}. Thus the expected allocations and payments
of a TIE mechanism form a deterministic truthful fractional mechanism. Moreover,
each machine's load under this mechanism is its expected load under $R$,
which is at most the expected makespan:
\begin{equation}\label{eq:mean-rule}
 \Mech{x}{t}\leq\E\,\Mech{R}{t}.
\end{equation}
The same reduction applies to randomized fractional mechanisms that are
truthful in expectation, with $x_{ij}(t)$ now the expected fraction of
task $j$ assigned to $i$. Their expected allocations and payments again
form a deterministic truthful fractional mechanism satisfying~\eqref{eq:mean-rule}.

A randomized mechanism is \emph{universally truthful} if it is a
probability distribution over truthful deterministic mechanisms. Every
universally truthful mechanism is TIE, so the reduction applies to it as
well.

\subsection{Benchmarks and approximation}
\label{sec:approx}

We write $\Opt{t}$ for the optimal \emph{integral} makespan, the minimum
of $\max_i t_i\cdot x_i$ over all integral allocations $x$, and
$\Optf{t}\leq\Opt{t}$ for the optimal fractional makespan, the same
minimum over all fractional allocations. We always distinguish these two
benchmarks. The performance of a randomized integral mechanism is its
expected makespan, and its benchmark is $\Opt{t}$. The performance of a
fractional mechanism is its makespan $\Mech{x}{t}$, and its benchmark is
$\Optf{t}$ unless stated otherwise. The approximation ratio of a mechanism on input $t$
is the ratio of its performance to its benchmark, and a mechanism is
\emph{$\rho$-approximate} if this ratio is at most $\rho$ for every input with
positive benchmark. (Both benchmarks vanish exactly when every task has a
zero-cost machine; the mechanisms of \cref{sec:upperbound} have makespan
zero at such inputs.)
For fixed $n$, a \emph{mechanism family} $R=(R_m)_{m\ge1}$ supplies
one mechanism for every finite number of tasks; it is
$\rho$-approximate if every $R_m$ is. The mechanisms for different $m$
may be arbitrary; there is no consistency requirement between task
counts.

Because of~\eqref{eq:mean-rule}, it suffices to prove our
lower bound for deterministic truthful fractional mechanisms, \emph{measured against
the integral optimum}. Thus we assume throughout the lower-bound proof
that
\begin{equation}\label{eq:integral-bound}
 \Mech{x}{t}\leq\rho\cdot\Opt{t}\qquad\text{for every input $t$ with
 $\Opt{t}>0$.}
\end{equation}
We need no assumption at inputs with zero optimum; every finite padded
instance to which we apply~\eqref{eq:integral-bound} has positive optimum.

\paragraph{Conventions.}
Throughout the lower-bound proof, \emph{fractional mechanism} means a
deterministic fractional mechanism. All inputs are finite matrices with
nonnegative real entries. We assume no regularity (such as
continuity) of the mechanism as a function of the reports.

\subsection{Weak monotonicity and disutilities}
\label{sec:wmon}

\begin{definition}[Disutility]
\label{def:disutility}
Fix a machine $i$ in a truthful mechanism and the values
$t_{-i}$ of the other machines. The \emph{disutility
function} of machine $i$ is
\[
 V_i(a;t_{-i})
 =a\cdot x_i(a,t_{-i})-P_i(a,t_{-i})+P_i(0,t_{-i}),
\]
and its value $V_i(a;t_{-i})$ is the \emph{disutility of machine $i$}.
We call $a$ the machine's \emph{own values} and $t_{-i}$ the
\emph{opposing values}.
Thus $V_i(a;t_{-i})$ is the decrease in the agent's truthful utility
when its type changes from $0$ to $a$, and $V_i(0;t_{-i})=0$; we always
use this normalization. The disutility is the machine's processing cost
minus its payment, shifted so that it vanishes when the agent's type
is~$0$; we reserve the word \emph{cost} for processing times.
When the machine and the values of the other machines
are fixed, we write simply $V(a)$.
\end{definition}

Fix a machine and the values of the other machines, and suppress them
in the notation for both $V$ and $x$. Applying~\eqref{eq:tie} with true
type $a$ and report $b$, and then with the roles exchanged, gives

\begin{equation}\label{eq:ic}
 x(b)\cdot(b-a)\;\leq\;V(b)-V(a)\;\leq\;x(a)\cdot(b-a).
\end{equation}
We refer to~\eqref{eq:ic} as the \emph{incentive inequalities}. They imply
\emph{weak monotonicity (WMON)},
 \[ \bigl(x(b)-x(a)\bigr)\cdot(b-a)\leq0.\]
In particular, increasing a machine's value on a single task cannot
increase the fraction of that task allocated to it. These are the
standard consequences of truthfulness: the incentive inequalities
express the convex-analytic characterization of Rochet~\cite{Roc87}, and
WMON is the monotonicity condition studied by~\cite{BCL+06,SaksYu2005}.

These inequalities also give the multi-dimensional form of Myerson's
payment identity~\cite{Mye81}. The function $V$ is concave, nondecreasing, and $1$-Lipschitz
(in the $\ell_1$ norm), with $0\leq V(a)\leq\sum_j a_j$. Along a single
coordinate, the change in $V$ is the area under the allocation curve:
\begin{equation}\label{eq:area}
 V(a+u\,\one_j)-V(a)=\int_0^u x_j(a+r\,\one_j)\,dr,
\end{equation}
with $\one_j$ the $j$-th unit vector. (Split $[0,u]$ into small
intervals and apply~\eqref{eq:ic} to each.) Equivalently,
$x_j=\partial V/\partial a_j$ wherever this partial derivative exists,
which is almost everywhere since $V$ is concave. These facts hold for
arbitrary tie-breaking. Along a coordinate line,~\eqref{eq:area} is the
envelope identity for one-parameter agents (see Archer and
Tardos~\cite{AT01}); Milgrom and Segal~\cite{MS02} give the general
envelope theorems in integral and derivative form.

All these identities concern a single machine. The two-task separation
of \cref{lem:two}, which drives our lower bound, is of a different
nature: it combines the incentive inequalities of \emph{both} machines
that compete for the same two tasks, and has no counterpart for a single
agent.

\paragraph{Restrictions to sets of tasks.}
Fix all machines' values on tasks outside a set $J$. The restriction of
a truthful mechanism to $J$ is again truthful: subtract each machine's
processing cost on the fixed tasks from its payment, so that its utility
is unchanged. Those processing costs can vary because the allocations of
the fixed tasks can vary. We call this restricted mechanism the
\emph{$J$-slice}; for $J=\{e,f\}$, the \emph{$(e,f)$-slice}.
Its disutility function is zero when the machine's values on $J$ are all zero.
In particular, if $J'\subseteq J$ and the remaining coordinates of $J$
are fixed at $z$, the disutility functions satisfy
\begin{equation}\label{eq:projection}
 V_{J'}(a;z)=V_J(a,z)-V_J(0_{J'},z).
\end{equation}

\paragraph{Additivity and symmetry.}
Fix a machine and all values except its own values on a set $D$ of
tasks, and let $V$ be its disutility function on $\R_{\geq0}^D$. For
distinct $e,f\in D$, the \emph{rectangle differences} of $V$ in the
coordinates $e,f$ are
\[
 V(a+u\one_e+v\one_f)-V(a+u\one_e)-V(a+v\one_f)+V(a),
 \qquad u,v\geq0,\ a_e=a_f=0.
\]
The function $V$ is \emph{additive} if
\[
 V(a)-V(0_D)=\sum_{e\in D}\bigl[V(a_e\one_e)-V(0_D)\bigr]
 \qquad\text{for every }a\in\R_{\geq0}^D;
\]
that is, the increment from raising one value does not depend on the
other values in $D$. Telescoping over the coordinates of $D$ shows that
$V$ is additive exactly when all its rectangle differences vanish. The
function $V$ is \emph{symmetric} if it is unchanged by every permutation
of the coordinates in $D$. When we say that $V$ is additive or symmetric
\emph{within} a subset of $D$, we apply these definitions to that
subset, with the values on the rest of $D$ fixed.
\subsection{Graph instances}
\label{sec:graphs}

Our lower-bound instances are described by multigraphs on the machines.
A task that only two machines $u,v$ can process at reasonable cost is an
\emph{edge} between them. Its \emph{endpoints} $u,v$ are its
\emph{eligible machines}, and all other machines have large finite
values on it. A task that only one machine $i$ can process is a
\emph{self-loop} at $i$, and $i$ is its \emph{owner}. Edges with the
same endpoints are \emph{parallel}. A \emph{multi-clique} has the same
number of parallel edges between every pair of machines. A \emph{star}
has one edge from a \emph{root} to each of the other machines, its
\emph{leaves}. The \emph{multi-star} of a root consists of all edges
incident to it.

\section{Lower Bound}
\label{sec:lb}

We prove the lower bound for deterministic truthful fractional mechanisms.
The reduction in \cref{sec:prelim} then gives the same lower bound for
randomized truthful-in-expectation mechanisms, with either integral or
fractional allocations.

\begin{theorem}\label{thm:lower}
Every deterministic truthful fractional mechanism for scheduling $n$ unrelated machines
has approximation ratio at least $(n+1)/2$, even against the optimal
\emph{integral} makespan, when arbitrary finite numbers of tasks are
allowed. More precisely, for every family $(R_m)_{m\geq1}$ of such
mechanisms and every $\delta>0$, there are $m$ and an input $t$ with $m$
tasks and $\Opt{t}>0$ such that $\Mech{R_m}{t}>((n+1)/2-\delta)\Opt{t}$.
\end{theorem}

\begin{corollary}\label{cor:tie}
Every randomized truthful-in-expectation mechanism, with integral or
fractional allocations, for scheduling $n$ unrelated machines has expected
approximation ratio at least $(n+1)/2$ against the optimal integral
makespan. In particular, this holds for every universally truthful
mechanism. Every deterministic truthful fractional mechanism has ratio
at least $(n+1)/2$ against the optimal fractional makespan.
\end{corollary}

\begin{proof}
The first claim follows from \cref{thm:lower} applied to the mechanism
formed from the expected allocations and payments,
by~\eqref{eq:mean-rule}; the last from $\Optf{t}\leq\Opt{t}$.
\end{proof}

\subsection{Technical overview}
\label{sec:overview}

Fix a family $(R_m)_{m\geq 1}$ of truthful fractional mechanisms, one for each number $m$ of
tasks, satisfying~\eqref{eq:integral-bound}, and assume $n\geq2$.

\paragraph{The hard instance.}
The hard instance is built on a \emph{multi-clique}: $\ell$ parallel
tasks between every pair of machines, plus one self-loop at each
machine. We call these the \emph{core tasks}; a padding construction
later adds further tasks. Most of the argument concerns the
\emph{multi-star} of a chosen root $i$, that is, all core tasks between
$i$ and the other machines. They form $k$ classes of $\ell$ tasks, one
class for each other machine.

\begin{figure}[ht]
\centering
\begin{tikzpicture}[
 vertex/.style={circle,draw,fill=blue!8,minimum size=7mm,inner sep=1pt},
 edge/.style={thick,blue!65!black},
 lab/.style={font=\small,fill=white,inner sep=1pt}]
 \node[vertex] (r) at (0,0.5) {$i$};
 \node[vertex] (a) at (-3,-1.6) {$j_1$};
 \node[vertex] (b) at (-1,-1.6) {$j_2$};
 \node[vertex] (d) at (3,-1.6) {$j_k$};
 \node at (1,-1.6) {$\cdots$};
 \foreach \l in {a,b,d}{
  \draw[edge] (r) -- (\l) node[lab,pos=.28]{$c$}
                           node[lab,pos=.78]{$\lambda$};}
 \draw[thick,orange!75!black] (r) to[out=55,in=125,looseness=9]
       node[lab,above,text=black] {$p_i$: value $1$} (r);
\end{tikzpicture}
\caption{The selected star within the multi-clique. Each selected task has value $c<1$ at the root and
$\lambda\geq1$ at its leaf; the self-loop $p_i$ has value $1$ at its owner, the root.
Assigning the star tasks to their leaves gives makespan $\lambda$, but the
root keeps about half of them. Under the continuity assumption of
\cref{sec:continuous} we use $\lambda=1$; in general,
$\lambda\in(1,1+\varepsilon)$. All other core tasks have a zero-cost
endpoint or owner.}
\label{fig:star}
\end{figure}
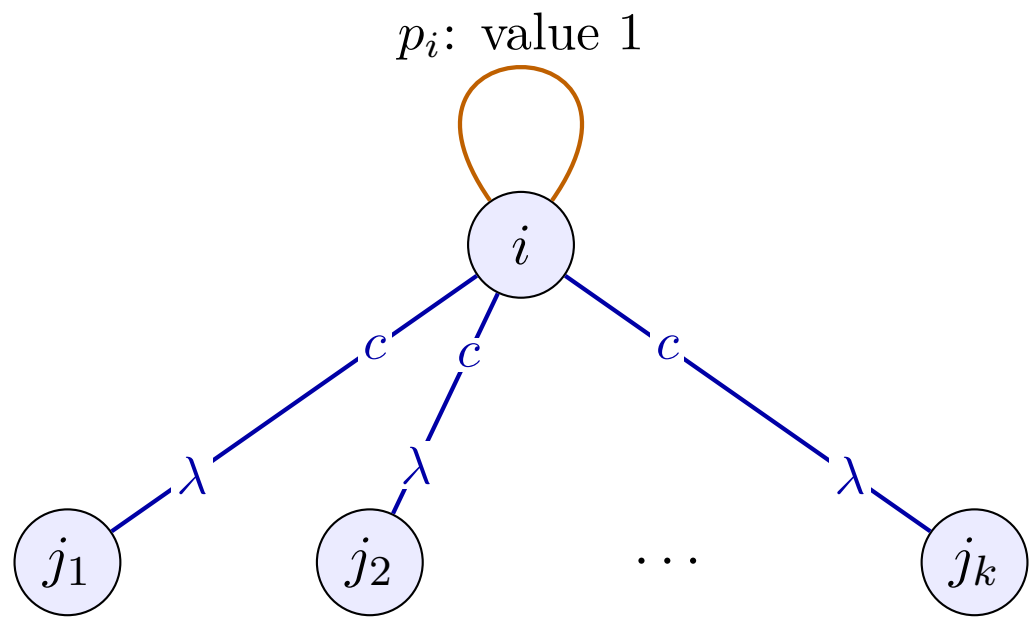

In the final input, we select one task $e_j$ from each class; these
tasks form a star $S$ inside the multi-star (\cref{fig:star}). The root's value on each
selected task is $c<1$ and its leaf's value is $\lambda\geq1$, both close
to $1$. The root
also owns a self-loop $p_i$ of value $1$. All remaining core tasks have a
zero-cost machine. Assigning every star task to its leaf and $p_i$ to the
root gives an integral schedule of makespan $\lambda$. We show that the
root receives a total fraction of at least $k/2-o(1)$ of the star tasks,
so its load is at least $(n+1)/2-o(1)$. The self-loop does not
increase the makespan of the benchmark schedule, but it adds one to the
root's load.

\paragraph{The task-independent case.}
For a \emph{task-independent} mechanism the argument is
short~\cite{CKK10,LY08}. Let $w_{ij}$ be the fraction of a task between
$i$ and $j$ that $i$ receives when both endpoint values are~$1$. Then
$w_{ij}+w_{ji}=1$, so the weights $w_{ij}$ sum to $\binom n2$ over all
ordered pairs, and some root $i$ has $\sum_{j\neq i}w_{ij}\geq k/2$.
With task independence, the root keeps at least $w_{ij}$ of its star task
towards $j$ whatever happens on the other tasks, so it keeps at least
$k/2$ of the star. A general mechanism, however, may couple the
allocations of different tasks arbitrarily. We use the parallel tasks of
the multi-star to recover enough task independence: the root's disutility
function becomes additive over the tasks of each class. Three ideas make
this work.

\paragraph{Idea 1: reservoirs give zero-type irrelevance (\cref{sec:padding}).}
We add many \emph{reservoir} tasks, each with a zero-cost endpoint, to
every pair of machines. We then average the mechanism over task
relabellings and over the numbers of reservoir tasks of each kind.
(Averaging over relabellings is analogous to the symmetrization of
Ashlagi, Dobzinski, and Lavi~\cite{ADL12}, who average over machine
permutations; in both cases truthfulness and the approximation guarantee
are preserved.) Since the approximation ratio is bounded independently
of the number of tasks, this yields in the limit a truthful mechanism on
a fixed finite graph with two properties. First, \emph{unique zeros}: a task with a unique zero-cost
endpoint goes wholly to that endpoint. Second, \emph{zero-type irrelevance}:
changing a task's endpoint values, while retaining at least one zero
endpoint value, leaves the allocation of every other task unchanged.
Zero-type irrelevance is the weak form of task independence that the
next idea turns into additivity. Both properties come from comparing a
core task with its identical reservoir copies; the paragraph after
\cref{lem:padding} explains how.

\paragraph{Idea 2: two-task separation by Laplace transforms
(\cref{sec:two}).}
Zero-type irrelevance pins down a machine's disutility on the
coordinate axes: if its value on one of two parallel tasks is zero, its
disutility does not depend on the other machine's value on that task. We show
that this, together with truthfulness of \emph{both} machines, forces the
disutility to be additive over the two tasks. Let $A$ and $B$ be the
differences between each machine's two-task disutility and the sum of its
one-task disutilities. By the observation above, $A$ vanishes when the
first machine's value on either task is zero, and $B$ vanishes when the
second machine's value on either task is zero. Truthfulness of both
machines gives the conservation law
$\partial_tA=-\partial_sB$. Integrating it against exponential weights
gives two independent linear relations between the Laplace transforms of
$A$ and $B$, so both transforms vanish, and hence $A=0$.

\paragraph{Idea 3: an exact identity for a fixed star (\cref{sec:retention}).}
Additivity holds within each class of parallel edges, but need not hold
across different leaves. Suppose the root's values on a class are all
zero. By additivity and symmetry, increasing one of these values to
$v\geq0$ changes its disutility by exactly $1/\ell$ of the change obtained by
increasing all values in the class to $v$. Applying this identity one
class at a time expresses the root's disutility as a weighted sum of its disutilities
at inputs where each class has root values all $0$ or all $v$.
This shows that the root keeps at least a $(1-1/\ell)^{k-1}$ fraction of
the total weight $\sum_{j\neq i}w_{ij}$, and this factor tends to $1$ as
$\ell\to\infty$.

\paragraph{Continuity first.}
For readability, we first give the complete argument assuming that every
machine's disutility function is jointly continuous, i.e., continuous on the values of all machines (\cref{sec:continuous},
ending with \cref{thm:continuous}). Under this assumption, two-task
separation holds at every positive opposing value, and we can take the
common leaf value to be $\lambda=1$.

The limiting construction in Idea~1 need not produce continuous disutility
functions. In \cref{sec:remove-continuity}, we show that the additivity
needed in Idea~3 can nevertheless be obtained. The additional work is
to justify the Laplace-transform argument without continuity and to
find a common leaf value at which the required additivity holds. We
can no longer simply set $\lambda=1$, but we can choose $\lambda>1$
arbitrarily close to $1$. The choice of the root, the estimate for its
allocation of the selected star, and the contribution of the
self-loop then proceed as in the continuous case, yielding the same lower
bound. The common leaf value $\lambda$ plays the role of the
\emph{continuity requirement} of~\cite{CKK26}, which is enforced there by
randomly perturbing the processing times.

\subsection{Core tasks and the padding lemma}
\label{sec:graph}
\label{sec:padding}

We use the graph instances of \cref{sec:graphs}, with vertex set $N$
and edge set $E:=M$; the self-loop at machine $i$ is denoted $p_i$. In
the finite scheduling inputs used below, the large finite processing
times of ineligible machines are specified in \cref{app:finite}.

\begin{definition}[Eligibility]\label{def:eligibility}
An allocation is \emph{eligible} if every edge is allocated only to its
endpoints and every self-loop only to its owner. An allocation
function is eligible if all its allocations are eligible.
\end{definition}

\begin{definition}[Core tasks and graph inputs]\label{def:core}
For an integer $\ell\geq2$, let $\mathcal{MC}_\ell$ consist of $\ell$ tasks between
each pair of machines and one self-loop $p_i$ at each machine. These
are the \emph{core tasks}. Write $Q_{\{u,v\}}$ for the class of tasks
between $u$ and $v$; tasks in the same class are \emph{parallel}.
A \emph{graph input} specifies the two endpoint values of every edge of
$\mathcal{MC}_\ell$ and the owner value of every self-loop, all
nonnegative, and no other values; thus each machine's values are those
on the core tasks of which it is an endpoint or the owner.
For a machine $i$, the union of the classes $Q_{\{i,j\}}$, $j\ne i$,
is the \emph{multi-star with root $i$}.
\end{definition}

Write $C(t)$ for the optimal integral makespan of graph input $t$ among
eligible allocations. This function is continuous, since it is the
minimum of the makespans of finitely many eligible integral allocations.

\begin{definition}[Zero-type irrelevance]\label{def:zero}
The \emph{type} of an edge is the pair of its two endpoint values. An
edge has a \emph{zero type} if at least one of these values is zero. An allocation function satisfies \emph{zero-type irrelevance} if changing
one edge from a zero type to another zero type, with all other values
fixed, leaves the allocation of every other task (including the
self-loops) unchanged.
\end{definition}

For example, under zero-type irrelevance, the allocation of the other tasks
does not change when an edge of type $(0,5)$ becomes $(0,0)$ or $(3,0)$.
The allocation of the changed edge itself may change.

\begin{definition}[Reservoir tasks and padded instances]\label{def:padded}
A \emph{reservoir task} is an additional task between a pair of machines
whose endpoint values are fixed at a zero type. A \emph{padded instance}
consists of the core tasks, finitely many reservoir tasks, and one
additional self-loop with positive owner value. All values in a
padded instance, including those at ineligible machines, are finite.
The numbers and values of the added tasks are specified in
\cref{app:finite}.
\end{definition}

The following lemma reduces the lower-bound problem to mechanisms with
additional structure. From a family of truthful fractional mechanisms
with finite approximation ratio, it constructs a truthful mechanism on
the core multi-clique whose
worst-case approximation ratio is no greater than that of the original
family. The constructed mechanism has structural properties that
facilitate the proof of the lower bound.
A lower bound for mechanisms with these properties therefore gives a
lower bound for the original family.

\Needspace{12\baselineskip}
\begin{lemma}[Padding]\label{lem:padding}
Let $R=(R_m)_{m\geq 1}$ be a family of deterministic truthful fractional
mechanisms on $n$ machines, one for every finite number of tasks, with
finite worst-case approximation ratio $\rho$ against the integral
optimum.
For every $\ell\geq2$ there is a truthful fractional mechanism $(x,P)$
on the core tasks of $\mathcal{MC}_\ell$, defined at all nonnegative real
graph inputs, whose approximation ratio against the eligible integral
optimum is no greater than that of $R$. More precisely,
\begin{equation}\label{eq:padding-load}
 \Mech{x}{t}\leq\rho\,C(t)\qquad\text{for every graph input $t$,}
\end{equation}
including inputs with $C(t)=0$. The mechanism also satisfies:
\begin{enumerate}[label=(\roman*),itemsep=2pt]
\item \emph{(Eligibility.)} Edges are allocated only to their endpoints,
 and self-loops only to their owners.
\item \emph{(Unique zeros.)} Every edge with a unique zero-cost endpoint
 is allocated wholly to that endpoint.
\item \emph{(Zero-type irrelevance.)} The allocation rule satisfies zero-type irrelevance (\cref{def:zero}).
\item \emph{(Symmetry.)} Relabelling tasks within any class $Q_{\{i,j\}}$
 relabels the allocation in the same way. Each machine's disutility is
 unchanged by the corresponding relabelling of the input.
\end{enumerate}
\end{lemma}

Consequently, if $(x,P)$ has approximation ratio greater than $\rho'$,
the original family also exceeds $\rho'$ on some finite input with
positive optimum. \Cref{app:witness} also gives a direct extraction
of such an input from the construction.

The proof of the lemma is in \cref{app:padding}. We give here the core
intuition.

\paragraph{Proof idea.}
The idea is to embed each graph input in larger instances, apply the given mechanisms there, and average. Fix an integer $K$. We pad every graph input with finitely many reservoir tasks: for each pair of machines and each zero type in a finite list $T_K$, between $K$ and $2K$ tasks of that type, together with one self-loop of small owner value that makes the optimum positive. Reservoir tasks can be scheduled at zero cost, so padding leaves the optimum essentially unchanged, while the approximation guarantee bounds every machine's load. We then average the allocations and payments over all relabellings of parallel tasks and over all numbers of reservoir tasks between $K$ and $2K$, keeping only the core tasks. The average remains truthful and satisfies the same load bound, since both are linear in allocations and payments. It is also symmetric, so identical tasks receive identical fractions. Thus, when a core task has a unique zero-cost endpoint, its many identical reservoir copies force its allocation to the positive-cost endpoint to vanish in the limit. The average is also insensitive to the numbers of reservoir tasks: changing one of them by one changes the average by $O(1/K)$. As $K\to\infty$, these facts give the properties of the lemma.

The least obvious property is zero-type irrelevance. Suppose a core edge
of zero type $\tau$ changes to another zero type $\sigma$. If at the
same time one reservoir task of the same class changes from $\sigma$ to
$\tau$, the multiset of types in every class and the total number of
tasks are preserved, so by symmetry every other core task receives
exactly the same allocation. The change of the reservoir task adds one
reservoir task of type $\tau$ and removes one of type $\sigma$, so by
insensitivity it changes the average by only $O(1/K)$.

Letting $K\to\infty$ along a subsequence gives a truthful mechanism at
rational inputs that satisfies the load bound and in which eligibility,
unique zeros and zero-type irrelevance hold exactly; rounding extends it
to real inputs. Conversely, a strict violation by the limit mechanism
persists at a rational input and some finite $K$, and hence in one term
of the finite averages, that is, for some $R_m$ at a padded instance.

\begin{definition}[Limit mechanism]\label{def:limit-mechanism}
For a fixed $\ell\geq2$, we call a mechanism obtained by the
construction in \cref{lem:padding} a \emph{limit mechanism}
on $\mathcal{MC}_\ell$.
\end{definition}

\subsection{The proof under continuity}
\label{sec:continuous}

In this subsection we prove the lower bound under the additional
assumption that every machine's disutility function is jointly continuous.

\begin{theorem}[The continuous case]\label{thm:continuous}
Suppose that, for every $\ell\geq2$, there is a truthful fractional
mechanism on $\mathcal{MC}_\ell$ satisfying properties~(i)--(iv) of
\cref{lem:padding} and the bound~\eqref{eq:padding-load} with the same
constant $\rho$, and every machine's
disutility function is jointly continuous on all graph values. Then $\rho\geq(n+1)/2$.
\end{theorem}

The assumption concerns disutility functions, not allocations, which may
still be discontinuous. It also passes to slices: by~\eqref{eq:projection},
every restricted disutility function is jointly continuous. The padding lemma
does not guarantee continuity, even for a deterministic  mechanism
for one task: if it gives the task to the root exactly when $t<\psi(s)$,
for a nondecreasing threshold function $\psi$, then it has disutility function
$V(t;s)=\min\{t,\psi(s)\}$. At any jump of $\psi$, this disutility jumps for
sufficiently large $t$, and the limit in \cref{lem:padding} gives no
control over such jumps. We remove the continuity assumption in
\cref{sec:remove-continuity}.

The proof of \cref{thm:continuous} has two ingredients, two-task
separation (\cref{sec:two}) and fixed-star retention
(\cref{sec:retention}), which are combined in \cref{sec:continuous-proof}.

\Needspace{26\baselineskip}
\subsubsection{Two-task separation}
\label{sec:two}
\label{sec:continuous-separation}

We state two-task separation once, in both forms needed. Part~(a) is used
in the continuous case and proved here; part~(b) is used in
\cref{sec:remove-continuity} and proved in \cref{app:two-measurable}.

\begin{lemma}[Two-task separation]\label{lem:two}
Let $(x,P)$ be a limit mechanism on $\mathcal{MC}_\ell$.
Choose two tasks $e,f$ with the same endpoints $i,j$, and
fix all values on the other tasks.

Write $t=(t_1,t_2)$ for machine $i$'s values on $e$ and $f$, and
$s=(s_1,s_2)$ for machine $j$'s values on them. Write $V_i(t;s)$
and $V_j(s;t)$ for the two machines' disutility functions in the
$(e,f)$-slice (\cref{sec:wmon}).

The identity
\begin{equation}\label{eq:two-separation}
 V_i(t_1,t_2;s)
 =V_i(t_1,0;s)+V_i(0,t_2;s)
 \qquad\text{for every }t\in\R_{\geq0}^2
\end{equation}
holds:
\begin{enumerate}[label=(\alph*),itemsep=2pt]
\item for every $s\in(0,\infty)^2$, if $V_i$ and $V_j$
 are jointly continuous on the four values;
\item for almost every $s\in(0,\infty)^2$, without any
 continuity assumption.
\end{enumerate}
The corresponding statements hold for $V_j$, with the
roles of the two machines exchanged.
\end{lemma}

The proof uses only truthfulness, eligibility, and
zero-type irrelevance. Consequently, the lemma also holds
for any mechanism satisfying properties (i)--(iv) of
\cref{lem:padding}.

\paragraph{Interpretation.}
For the opposing values covered by the lemma, machine $i$'s
allocation of each task is independent of its own value for the other
task, wherever the corresponding one-task disutility function is
differentiable. Tie-breaking may still depend on the other value; for
example, $V_i(t;s)=\min\{t_1,s_1\}+\min\{t_2,s_2\}$ is additive and
continuous, but at $t_1=s_1$ a truthful mechanism may split task~$1$
according to the values on task~$2$. Such ties are harmless: the
allocation bound of \cref{lem:retention} uses only additivity and
symmetry of the disutility function together with the incentive
inequalities, which hold at every report.

The proof follows Idea~2 of \cref{sec:overview}: Step~1 identifies the
disutilities on the coordinate axes, Step~2 introduces the differences
$A$ and $B$, Step~3 shows that all their Laplace transforms vanish, and
Steps~4 and~5 conclude.

\begin{proof}[Proof of part~(a)]
Let $t=(t_1,t_2)$ and $s=(s_1,s_2)$ be the values of the two machines,
and, for $h=1,2$, let $x_h$ and $1-x_h$ be the fractions of task $h$
allocated to machines $i$ and $j$. Their disutility functions $V_i(t;s)$ and
$V_j(s;t)$ vanish at their respective own zero values ($V_i(0;s)=V_j(0;t)=0$). They are
concave and $1$-Lipschitz in their own values, and satisfy
\[
 0\leq V_i(t;s)\leq t_1+t_2,
 \qquad 0\leq V_j(s;t)\leq s_1+s_2.
\]
By the area formula~\eqref{eq:area}, their own partial derivatives are the corresponding
allocations wherever these derivatives exist.

\emph{Step 1: the one-task parts.}
For $h=1,2$, let $V_{i,h}(t_h;s_h)$ and $V_{j,h}(s_h;t_h)$ be the two
machines' disutility functions when both machines' values on the other task,
$3-h$, are zero.
Suppose that machine $i$ has value zero on task~$2$. This task has a
zero type for every $s_2$. Zero-type irrelevance therefore lets us
change $s_2$ to zero without changing the allocation of task~$1$,
at every value of $t_1$. Thus the entire allocation curve for task~$1$
is unchanged. Since both disutility functions $V_i$ and $V_{i,1}$ are normalized to zero at
$t_1=0$, the area formula gives
\[
 \begin{aligned}
 V_i(t_1,0;s_1,s_2)
 &=\int_0^{t_1}x_1(r,0;s_1,s_2)\,dr\\
 &=\int_0^{t_1}x_1(r,0;s_1,0)\,dr\\
 &=V_{i,1}(t_1;s_1).
 \end{aligned}
\]
Exchanging the two tasks gives the other \emph{axis identity}:
\begin{equation}\label{eq:axes}
 V_i(t_1,0;s)=V_{i,1}(t_1;s_1),\qquad
 V_i(0,t_2;s)=V_{i,2}(t_2;s_2).
\end{equation}
The corresponding identities hold for $V_j$. 

Next we use truthfulness of \emph{both} machines. By the area formula,
each machine's fraction of a task is the derivative of its own disutility
function with respect to its value on that task. Since the two fractions
of a task sum to one, the two derivatives sum to one. The same holds for
the one-task disutility functions. Each identity holds wherever the
derivatives involved exist. For fixed opposing values, the own
derivatives exist almost everywhere, since the disutilities are Lipschitz in
their own values. The disutilities are jointly continuous, hence measurable, so
by Fubini's theorem the identities hold almost everywhere in all four
values. For $h=1,2$, they are
\begin{equation}\label{eq:complement}
 \partial_{t_h}V_i(t;s)+\partial_{s_h}V_j(s;t)=1,\qquad
 \partial_{t_h}V_{i,h}(t_h;s_h)+\partial_{s_h}V_{j,h}(s_h;t_h)=1.
\end{equation}

\emph{Step 2: the disutility differences.}
Subtract the one-task parts:
\[
 \begin{split}
 A(t,s)&=V_i(t;s)-V_{i,1}(t_1;s_1)-V_{i,2}(t_2;s_2),\\
 B(t,s)&=V_j(s;t)-V_{j,1}(s_1;t_1)-V_{j,2}(s_2;t_2).
 \end{split}
\]
Our aim is to show that $A=0$. By~\eqref{eq:axes}, $A$ vanishes when
$t_1=0$ or $t_2=0$, and $B$ vanishes when $s_1=0$ or $s_2=0$.
We introduce both differences because~\eqref{eq:complement} relates
their derivatives. Given that
\[
 \begin{split}
 \partial_{t_h}A(t,s)&=\partial_{t_h}V_i(t;s)-\partial_{t_h}V_{i,h}(t_h;s_h),\\
 \partial_{s_h}B(t,s)&=\partial_{s_h}V_j(s;t)-\partial_{s_h}V_{j,h}(s_h;t_h),
 \end{split}
\]
where the derivatives exist, subtracting the two equalities in \eqref{eq:complement} yields $\partial_{t_h}A+\partial_{s_h}B= 1-1 =0$ almost everywhere. Define for such $(t,s)$ 
\begin{equation}\label{eq:derivatives}
 g_h:=\partial_{t_h}A=-\partial_{s_h}B,
 \qquad |g_h|\leq1\qquad(h=1,2)
\end{equation}
 where the bound holds by \eqref{eq:ic}. We also have
$|A(t,s)|\leq t_1+t_2$ and $|B(t,s)|\leq s_1+s_2$.

\emph{Step 3: two equations for the transforms.}

We use Laplace transforms to convert the two derivative identities
into linear equations for the same two unknowns. A Laplace
transform integrates a function against a decaying exponential
weight. Since the derivative of this weight is a constant
multiple of itself, integration by parts replaces differentiation
by multiplication by the corresponding decay rate, provided
the boundary terms vanish (cf. Theorem 9.2 in \cite{Doetsch74}). Applying this to both identities
in~\eqref{eq:derivatives} gives two equations for the transforms
of $A$ and $B$. By choosing the rates so that these equations
are independent, we will show that both transforms vanish.

For positive rates $\alpha=(\alpha_1,\alpha_2)$ and
$\gamma=(\gamma_1,\gamma_2)$, write
\[
 \mathcal L_{\alpha,\gamma}(F)
 =\int_{\R_{\geq0}^4}F(t,s)
       e^{-\alpha\cdot t-\gamma\cdot s}\,dt\,ds.
\]
The growth bounds make these integrals absolutely convergent for
$A$, $B$, and $g_h$, so Fubini's theorem applies. Moreover, $A$ is
Lipschitz in each $t_h$, and $B$ is Lipschitz in each $s_h$.
Their restrictions to these coordinates are therefore absolutely
continuous, which justifies integration by parts even though their
derivatives need exist only almost everywhere.
Integrating by parts in $t_h$ gives (e.g., for $h=1$)

\[
\begin{split}
\mathcal L_{\alpha,\gamma}(A)
 &= \int_{\R_{\geq0}^4}A\cdot
       e^{-\alpha\cdot t-\gamma\cdot s}\,dt\,ds\\
       &=\int_{\R_{\geq0}^3} \left[A\cdot(-\alpha_1^{-1})\cdot
       e^{-\alpha\cdot t-\gamma\cdot s}\right]_{t_1=0}^{\infty}\,dt_2\,ds-\int_{\R_{\geq0}^4} \partial_{t_1}A\cdot (-\alpha_1^{-1})\cdot
       e^{-\alpha\cdot t-\gamma\cdot s}\,dt\,ds\\
   &=0 + \frac{1}{\alpha_1}\int_{\R_{\geq0}^4}g_1\cdot
       e^{-\alpha\cdot t-\gamma\cdot s}\,dt\,ds
       =\frac{1}{\alpha_1}\mathcal L_{\alpha,\gamma}(g_1).     
\end{split}
\]
The boundary term at zero vanishes because $A=0$ when $t_1=0$,
and the term at infinity vanishes because $A$ grows at most linearly
whereas the exponential weight decays.

Thus for $h=1,2$ we obtain $\alpha_h\mathcal L_{\alpha,\gamma}(A)
 =\mathcal L_{\alpha,\gamma}(g_h).$ Similarly, integration by parts in $s_h$ gives
$\gamma_h\mathcal L_{\alpha,\gamma}(B)
 =-\mathcal L_{\alpha,\gamma}(g_h)$, using $B=0$ when $s_h=0$
and the corresponding growth bound.
Thus
\[
 \alpha_h\mathcal L_{\alpha,\gamma}(A)
 =-\gamma_h\mathcal L_{\alpha,\gamma}(B)
 \qquad(h=1,2).
\]
For fixed rates, these are two linear equations for the same two
numbers $\mathcal L_{\alpha,\gamma}(A)$ and
$\mathcal L_{\alpha,\gamma}(B)$. Their determinant is
$\alpha_1\gamma_2-\alpha_2\gamma_1$, so both transforms vanish whenever
$\alpha_1/\gamma_1\neq\alpha_2/\gamma_2$.
Such choices of rates are dense among all positive rates.
The growth bounds imply continuity of the transforms in the positive
rates by dominated convergence, so the transforms vanish for every
choice of positive rates.

\emph{Step 4: vanishing almost everywhere.}
The functions $A$ and $B$ are measurable, satisfy the growth bounds
$|A(t,s)|\leq t_1+t_2$ and $|B(t,s)|\leq s_1+s_2$, and all their
Laplace transforms with positive rates vanish. By the uniqueness of the
Laplace transform,
$A=B=0$ almost everywhere (\cref{lem:laplace} in \cref{app:laplace}).\footnote{See also Theorem 5.1 in \cite{Doetsch74}.} No continuity of $A$ or $B$ is needed in
this step.

\emph{Step 5: the pointwise conclusion under joint continuity.}
For part~(a), $A$ is jointly continuous. If it were nonzero at some
$t\in\R_{\geq0}^2$ and $s\in(0,\infty)^2$, it would remain nonzero
on a relative neighbourhood of that point. Such a neighbourhood
has positive four-dimensional measure, contradicting Step~4.
Thus $A(t,s)=0$ for every such $t,s$. By~\eqref{eq:axes}, this is
exactly~\eqref{eq:two-separation}. The same argument with the machines
exchanged gives the corresponding assertion for $V_j$.

In part~(b), Step~4 will instead be applied to measurable versions
of the disutility functions. Fubini's theorem and continuity in the own
values will then give additivity for every own value outside one
null set of opposing values, as detailed in
\cref{app:two-measurable}.
\end{proof}

\subsubsection{Fixed-star retention}
\label{sec:retention}

The main result of this subsection is a lemma that combines the estimates for individual tasks into a bound for a star. It applies to a machine's disutility function that is additive within each class of coordinates, but need not be additive across classes. The high-level intuition is that even in the non-additive case, changing a single task has bounded total influence due to  the incentive inequalities \eqref{eq:ic}. The lemma does not require continuity, so we will also use it in the general case.

Consider a truthful fractional mechanism and fix a machine.
Focus on its values on a set $D$ of tasks, keeping all other
values fixed. Write $V(a)$ for its disutility and $x_e(a)$
for the fraction of task $e$ allocated to it when its values
on $D$ are $a$. Partition $D$ into $k$ classes $D_1,\ldots,D_k$, each of size
$\ell$. Choose one task $e_j\in D_j$ from each class and set
$S=\{e_1,\ldots,e_k\}$. Let $q_j$ be the fraction of task $e_j$ allocated to the machine
when it reports $1$ on that task alone. Also, for $0<u<1$, let $N_u$
be its {\em total allocation} of $S$ when it reports $u$ on {\em every}
task in $S$. Both reports are zero elsewhere in $D$. Equivalently,
\[
 q_j=x_{e_j}(\one_{e_j}),
 \qquad
 N_u=\sum_{j=1}^k x_{e_j}(u\one_S).
\]

\begin{lemma}[Fixed-star retention]\label{lem:retention}
Suppose that $V$ is additive and symmetric within each class
$D_j$, for every fixed choice of values on the other classes.
Then, for every $0<u<1$,
\[
 N_u\geq
 \left(1-\frac1\ell\right)^{k-1}\sum_{j=1}^k q_j.
\]
\end{lemma}

We remark that if $V$ were additive over {\em all} tasks in $S$,
we would immediately obtain $N_u\geq\sum_{j=1}^kq_j$.
Here we assume additivity only within each class, and the
factor $(1-1/\ell)^{k-1}$ is the price of interactions between
classes. Following Idea~3 of \cref{sec:overview}, the proof
expands $V(v\one_S)$ exactly over unions of whole classes, and then subtracts the expansion at $v=u$ from
that at $v=1$ and keeps only the single-class terms in the
resulting difference.

\begin{proof}

First we focus on a fixed single class $D_j$. Let $v\in\{u,1\}$ and let $t$ be a value vector that is zero on
each task of $D_j$. By additivity within $D_j$ and symmetry,
\[
\begin{aligned}
 V(t+v\one_{D_j})-V(t)
 &=\sum_{e\in D_j}\bigl[V(t+v\one_e)-V(t)\bigr]\\
 &=\ell\bigl[V(t+v\one_{e_j})-V(t)\bigr].
\end{aligned}
\]
Rearranging gives
\begin{equation}\label{eq:single-class}
 V(t+v\one_{e_j})
 =\left(1-\frac1\ell\right)V(t)
  +\frac1\ell\,V(t+v\one_{D_j}).
\end{equation}

We next apply \eqref{eq:single-class} successively to all classes. For $A\subseteq\{1,\ldots,k\}$ let
$D_A=\bigcup_{j\in A}D_j$. As an illustration, for $k=2$ \eqref{eq:single-class} yields the following formula for the disutility of $\one_{\{e_1,e_2\}}$ (with $v=1$). In the first step $\one_{e_1}$ plays the role of $t;$ after that we use \eqref{eq:single-class} setting $t=0$ and $t=\one_{D_2},$ respectively:

\[
 \begin{split}
 V(\one_{\{e_1,e_2\}})=
 V(\one_{e_1}+\one_{e_2})&= \left(1-\frac1\ell\right)V(\one_{e_1})+\frac1\ell\,V(\one_{e_1}+ \one_{D_2})\\
 &=\left(1-\frac1\ell\right)\left[\left(1-\frac1\ell\right)V(0)+\frac1\ell V(\one_{D_1})\right]\\
 &+\frac1\ell\left[\left(1-\frac1\ell\right)V(\one_{D_2})+\frac1\ell V(\one_{D_2}+\one_{D_1})\right]\\
 &=\left(1-\frac1\ell\right)^2 V(0)+\frac1\ell\left(1-\frac1\ell\right)(V(\one_{D_1})+V(\one_{D_2}))+\frac{1}{\ell^2}V(\one_{D_1\cup D_2}).
 \end{split}
\]
For general $k$, induction using~\eqref{eq:single-class}
gives the exact identity
\begin{equation}\label{eq:expansion}
 V(v\one_S)=\sum_{A\subseteq\{1,\ldots,k\}}
  \ell^{-|A|}\left(1-\frac1\ell\right)^{k-|A|}V(v\one_{D_A}).
\end{equation}
All coefficients are nonnegative and sum to one, by the binomial theorem.

Finally, subtracting the instance of~\eqref{eq:expansion}
with $v=u$ from the instance with $v=1$ gives
\[
 V(\one_S)-V(u\one_S)
 =\sum_{A\subseteq\{1,\ldots,k\}}
 \ell^{-|A|}\left(1-\frac1\ell\right)^{k-|A|}
 \bigl[V(\one_{D_A})-V(u\one_{D_A})\bigr].
\]
Since $V$ is nondecreasing, every term on the right is
nonnegative. We therefore retain only the terms with
$A=\{j\}$. For these,
additivity, symmetry and the left incentive inequality of~\eqref{eq:ic} give
\[
 V(\one_{D_j})-V(u\one_{D_j})
 =\ell\bigl(V(\one_{e_j})-V(u\one_{e_j})\bigr)
 \geq \ell(1-u)\,q_j .
\]
On the left side, the right incentive inequality of~\eqref{eq:ic} gives
$V(\one_S)-V(u\one_S)\leq(1-u)N_u$. Putting these bounds together, we obtain
\[
 (1-u)N_u\geq
 \sum_{j=1}^k\ell^{-1}\left(1-\frac1\ell\right)^{k-1}
 \ell(1-u)\,q_j,
\]
which proves the lemma.
\end{proof}

\subsubsection{Proof of \texorpdfstring{\cref{thm:continuous}}{The continuous-case theorem}}
\label{sec:continuous-proof}
\label{sec:star}

The proof consists of four parts: first we select the root,
then we establish additivity and symmetry within each class,
next we bound the root's allocation of the selected tasks,
and finally we add the self-loop.

\begin{proof}[Proof of \cref{thm:continuous}]
Fix $\ell\geq2$ and a mechanism as in \cref{thm:continuous}.

We first fix a machine, the \emph{root}, which would get at least half of a star of unit-value tasks had the mechanism been task independent. The main point of the proof is to show that the root keeps almost this share without assuming task independence.
For each unordered pair of machines $\{i,j\}$, choose one
core task $e_{ij}$ between them. Consider the input
$\mathcal I_{ij}$ in which both endpoint values of $e_{ij}$
are $1$, all endpoint values of the other core tasks
are $0$, and every self-loop has owner value $0$.

Let $w_{ij}$ and $w_{ji}$ be the fractions of $e_{ij}$ allocated to machines $i$ and $j$, respectively, on input $\mathcal I_{ij}$. By eligibility, $w_{ij}+w_{ji}=1$, so $\sum_i\sum_{j\ne i}w_{ij}=\binom n2$, and some machine $i$ satisfies
\[ \sum_{j\ne i}w_{ij}\geq\frac1n\binom n2=\frac k2. \]
We select such a machine as the root.

We next show that the root's disutility is additive and symmetric within each class. 
For each machine $j\ne i$, let $D_j=Q_{\{i,j\}}$ be the
$\ell$ core tasks between the root $i$ and $j$, and let
$D=\bigcup_{j\ne i}D_j$. Fix a common leaf value $\lambda\geq1$, and
give each machine $j\ne i$ value $\lambda$ on all tasks in $D_j$. The
continuous case uses only $\lambda=1$. We allow $\lambda\ge 1$, because the rest of the argument is reused in \cref{sec:remove-continuity}. Set both
endpoint values of every core edge not incident to $i$ to $0$, and every
self-loop's owner value to $0$.

Keep these values fixed and let the root's values
$t=(t_e)_{e\in D}$ vary. Let $V$ be the root's disutility
function.
Fix $j\ne i$ and two tasks $e,f\in D_j$. By~\eqref{eq:projection}, the
rectangle differences of $V$ in the coordinates $e,f$ are those of the
root's disutility in the $(e,f)$-slice. Hence, by \cref{lem:two}\,(a) at the
opposing values $(\lambda,\lambda)$, the root's disutility function is additive
in its values on $e$ and $f$, with all other values fixed arbitrarily. Consequently, the change in
disutility from increasing one task value is independent of
the values on the other tasks in $D_j$. Adding the tasks
one at a time gives
\[
 V(t)-V(z)
 =\sum_{e\in D_j}\bigl[V(z+t_e\one_e)-V(z)\bigr],
\]
where $z$ agrees with $t$ outside $D_j$ and is zero on $D_j$.
Thus $V$ is additive within $D_j$.

All values of machine $j$ on $D_j$ are equal to $\lambda$, so
relabelling the tasks of $D_j$ does not change the fixed values. By
property~(iv) of \cref{lem:padding}, $V$ is therefore symmetric in the
root's values on $D_j$. Hence $V$ satisfies the additivity
and symmetry assumptions of \cref{lem:retention}.

We now bound the fraction of the selected tasks that the root keeps.
For each $j\ne i$, let $e_j\in D_j$ be the task $e_{ij}$ chosen
above, and let
$S=\{e_j:j\ne i\}$. Keep the other machines' values fixed
as above, with common leaf value $\lambda\geq 1$.
Let $q_j$ be the fraction of $e_j$ allocated to the root
when its value is $1$ on $e_j$ and $0$ on all other
tasks in $D$.

At this input, $e_j$ has value $1$ at the root and $\lambda$ at
$j$, and every other task of $D$ has value $0$ at the root and $\lambda$
at its leaf. These other tasks have zero types; change each of them,
one at a time, to values $(0,0)$. Zero-type irrelevance leaves the
allocation of $e_j$ unchanged.

The resulting input differs from $\mathcal I_{ij}$ only in machine $j$'s value on $e_j$, which is
$\lambda$ instead of $1$. By truthfulness, increasing
this value cannot increase the fraction allocated to $j$.
Since the two endpoint fractions sum to $1$, the fraction
allocated to the root cannot decrease. Hence
\begin{equation}\label{eq:q}
 q_j\geq w_{ij},\qquad
 \sum_{j\ne i}q_j\geq\frac{k}{2},
\end{equation}
with $q_j=w_{ij}$ when $\lambda=1$.

For $0<u<1$, define $\mathcal I_u$ as follows. The root
has value $u$ on $S$ and $0$ on $D\setminus S$; each
leaf has value $\lambda$ on every task in its class.
All core edges not incident to the root have endpoint
values $(0,0)$, and all self-loops have owner value $0$.
Define
 \[ N_u:=\sum_{e\in S}x_{ie}(\mathcal I_u).\]
By \cref{lem:retention}, using the additivity and symmetry
established above,
\begin{equation}\label{eq:Nu}
 N_u\geq
 \left(1-\frac1\ell\right)^{k-1}\sum_{j\ne i}q_j
 \geq\frac{k}{2}\left(1-\frac1\ell\right)^{k-1}.
\end{equation}

Finally, we add a unit self-loop at the root.
Choose rational $0<c<u<1$. Obtain $\mathcal I^*$ from
$\mathcal I_u$ by changing the root's value on every task
in $S$ from $u$ to $c$, and its value on its self-loop
$p_i$ from $0$ to $1$. Keep every other value unchanged.

Let $Q_c=\sum_{e\in S}x_{ie}(\mathcal I^*)$ be the total fraction of $S$
allocated to the root in $\mathcal I^*$. The self-loop $p_i$ is
allocated wholly to the root at both inputs, by eligibility. The two
inputs differ only in the root's values, so WMON for the root gives
\[
 (c-u)(Q_c-N_u)+(1-0)(1-1)\leq0,
\]
so $Q_c\geq N_u$. The root processes all of $p_i$, at value $1$, and a
total fraction $Q_c$ of the tasks in $S$, each at value $c$. By~\eqref{eq:Nu},
its load in $\mathcal I^*$ is therefore at least
\begin{equation}\label{eq:rootload}
 1+cQ_c\geq1+\frac{ck}{2}\left(1-\frac1\ell\right)^{k-1}.
\end{equation}
On the other hand, assign the self-loop to the root, each selected
task to its leaf, and all other tasks to zero-cost endpoints or
owners. This schedule has makespan at most $\lambda$, so
$C(\mathcal I^*)\leq\lambda$. For the continuous case, now take $\lambda=1$.
The root's self-loop has value one and must be assigned to the root,
so $C(\mathcal I^*)=1$. The load bound~\eqref{eq:padding-load} gives
\begin{equation}\label{eq:continuous-final}
 \rho\geq1+\frac{c(n-1)}2\left(1-\frac1\ell\right)^{n-2}.
\end{equation}

By the assumptions of the theorem, such a mechanism exists for every
$\ell\geq2$, with the same $\rho$. Thus~\eqref{eq:continuous-final} holds
for every $\ell$ and every rational $0<c<1$. Letting $\ell\to\infty$ and then
$c\uparrow1$ gives $\rho\geq(n+1)/2$.
\end{proof}

\subsection{Removing the continuity assumption}
\label{sec:remove-continuity}
\label{sec:separation}

We now drop the continuity assumption. Fix a limit mechanism from
\cref{lem:padding}. The retention lemma and most of the proof of \cref{thm:continuous} above need no
continuity. Continuity was used only to show that the root's disutility is additive within each class, in two ways: to guarantee that the
disutilities are jointly measurable, so that the integrals in the proof of two-task
separation make sense, and to apply
the resulting identity at the prescribed opposing values $(\lambda,\lambda)$, which
might lie in an exceptional set of measure zero. Both uses are replaced
using one consequence of zero-type irrelevance: monotonicity in the
opposing values. It provides jointly measurable versions of the disutilities
(\cref{app:two-measurable}) and a common leaf value $\lambda$ arbitrarily
close to one at which the disutility is continuous.

\paragraph{Monotonicity in opposing values.}

\begin{lemma}\label{lem:opponent}
Fix a root and vary only its values $t$ on its incident edges and the
opposing endpoint values $s$ on those edges; fix all other values. Then
the root's disutility function $V(t;s)$ is nondecreasing in each
coordinate of $s$.
\end{lemma}

\begin{proof}
Fix an incident edge $e$. By the area formula~\eqref{eq:area}, we can
reach $t$ from $0$ by first raising the values on the other edges and
then the value on $e$:
\[
 V(t;s)=V(t_{-e},0;s)+\int_0^{t_e}x_{ie}(t_{-e},r;s)\,dr .
\]
In the first term, the root has value $0$ on $e$ throughout, so $e$ has a
zero type whatever $s_e$ is. By zero-type irrelevance, changing $s_e$
does not change the allocations integrated in the first term, so this
term does not depend on $s_e$. In the second term, WMON for the opposing
machine says that its fraction of $e$ cannot increase when $s_e$
increases. By eligibility, the root's fraction $x_{ie}$ is the
complement, so it cannot decrease.
\end{proof}

\paragraph{Two-task separation almost everywhere.}
\Cref{lem:two}\,(b) gives the general form of two-task separation. Its
proof in \cref{app:two-measurable} uses \cref{lem:opponent} to construct
jointly measurable versions of the disutilities that agree with the originals at every own
value, for almost every opposing value. The axis identities and the
complementary-gradient identities remain valid for these versions, so
the Laplace-transform computation applies unchanged.

\paragraph{A common leaf value.}
The exceptional set of opposing values in \cref{lem:two}\,(b) may
contain $(1,1)$, or any other prescribed pair $(\lambda,\lambda)$. We therefore choose a nearby common leaf value at which
the root's disutility function is continuous. The following lemma replaces the part of
\cref{sec:continuous-proof} that needs continuity.

\begin{lemma}\label{lem:lambda}
Fix a root $i$ and $\varepsilon>0$. Give every core edge not incident to
$i$ value zero at both endpoints, and every self-loop owner value zero. Let
$D_1,\ldots,D_k$ be the $\ell$-edge classes incident to $i$. There
is $\lambda\in(1,1+\varepsilon)$ with the following property. When every
opposing endpoint value in these classes equals $\lambda$, the root's
disutility function $V$ is additive within each $D_j$, for every fixed choice
of the root's values on the other classes, and it is symmetric under
permutations of the coordinates within each~$D_j$.
\end{lemma}

\begin{proof}
Write $V(t;s)$ for the root's disutility when its values on the $k\ell$
incident tasks are $t$ and the opposing endpoint values are $s$.
Its self-loop value remains zero.

\emph{Choice of $\lambda$.} For each rational root vector $t$, the
function $f_t(z)=V(t;z\one)$ of a single variable $z$ is nondecreasing by
\cref{lem:opponent}, so it has at most countably many discontinuities.
Choose $\lambda\in(1,1+\varepsilon)$ outside the union of these countably
many countable sets. For any $s$ with $\|s-\lambda\one\|_\infty<\eta$,
monotonicity gives
\[
 f_t(\lambda-\eta)\leq V(t;s)\leq f_t(\lambda+\eta),
\]
so $V(t;\cdot)$ is continuous at $s=\lambda\one$ for every rational $t$.

\emph{Additivity for a pair.} Fix two parallel edges $e,f$ in a class $D_j$,
rational root values on the other incident edges, and opposing values
$\lambda$ on those edges. The $(e,f)$-slice is a truthful two-task
mechanism with complementary allocations and zero-type irrelevance. By
\cref{lem:two}\,(b), for almost every opposing pair $(s_e,s_f)$ and all
rational $a,b\geq0$, the rectangle difference vanishes:
\begin{equation}\label{eq:rectangle}
 V(t+a\one_e+b\one_f)-V(t+a\one_e)-V(t+b\one_f)+V(t)=0,
\end{equation}
where $t_e=t_f=0$. (By~\eqref{eq:projection}, the rectangle difference of
the full disutility equals that of the $(e,f)$-slice.) There are
countably many choices of the rational data, so we can approach
$(\lambda,\lambda)$ through opposing pairs at which~\eqref{eq:rectangle}
holds for all of them. By continuity at $s=\lambda\one$,
\eqref{eq:rectangle} also holds when all opposing values equal
$\lambda$.

\emph{Whole classes.} Within a class $D_j$, telescoping over the
coordinates and using~\eqref{eq:rectangle} to move each increment to its
own axis shows that $V$ is additive at rational root values. Since $V$
is Lipschitz in the root's values, this extends to all root values,
including arbitrary fixed values on the other classes. Finally, the
fixed values are unchanged by permutations within $D_j$, so property~(iv)
of \cref{lem:padding} makes $V$ symmetric within that class.
\end{proof}

\paragraph{Completing the star argument.}
Choose the root and the tasks $e_j$ exactly as in the proof of \cref{thm:continuous}. Fix $\varepsilon>0$ and choose
$\lambda\in(1,1+\varepsilon)$ by \cref{lem:lambda} for this root. Use this common leaf value for the leaves in all inputs after the choice of the root.
\Cref{lem:lambda} supplies the additivity and symmetry of the root's
disutility function. The rest of the proof is the same: this gives~\eqref{eq:q} and
\eqref{eq:Nu}; we then get the same root-load bound~\eqref{eq:rootload}
and the upper bound $C(\mathcal I^*)\leq\lambda$ on the optimum.

\begin{proof}[Proof of \cref{thm:lower}]
For $n=1$ the bound is immediate. For $n\geq2$, the preceding construction
and the load bound of \cref{lem:padding} give, for every $\ell\geq2$,
rational $0<c<1$, and $\varepsilon>0$,
\[ \rho(1+\varepsilon)\geq\rho\lambda
 \geq1+\frac{c(n-1)}2\left(1-\frac1\ell\right)^{n-2}.\]
Let $\ell\to\infty$, then $c\uparrow1$ and $\varepsilon\downarrow0$.
It follows that $\rho\geq(n+1)/2$.

To make the finite-witness statement explicit, suppose that every finite
positive-optimum input satisfied a bound $\rho'<(n+1)/2$. Choose $\ell$,
rational $0<c<1$, and $\varepsilon>0$ so that
\[
 1+\frac{c(n-1)}2\left(1-\frac1\ell\right)^{n-2}
 >\rho'(1+\varepsilon).
\]
Apply the padding lemma to the original family, whose approximation
ratio is at most $\rho'$. Its limit mechanism inherits this bound and
admits the star just constructed, with $C\leq\lambda<1+\varepsilon$,
contradicting that bound. Hence some finite positive-optimum input
violates the bound $\rho'$. \Cref{app:witness} also gives the direct
transfer from a strict graph-input violation to an original finite
input. No consistency between mechanisms at different task counts is
used.
\end{proof}

\section{Upper Bound}
\label{sec:upperbound}

In this section, we propose a
universally truthful mechanism, which we call
\textsc{Exp-Bounded-Square}, that achieves an expected approximation
ratio of at most $(n+1)/2 + O(\sqrt{n \ln n}) $
(\Cref{cor:upper-bound}). 

In \Cref{sec:exp-bounded-square}, we formally define
\textsc{Exp-Bounded-Square} and derive the task assignment
probabilities as a function of the machines' bids.  Next,
\Cref{sec:universal-truthfulness} establishes the universal
truthfulness of the mechanism, and \Cref{sec:approximation}
presents the approximation guarantee.

\subsection{The Exp-Bounded-Square mechanism}
\label{sec:exp-bounded-square}
\textsc{Exp-Bounded-Square}, presented in
\Cref{alg:exp-bounded-square}, is an adaptation of the
truthful-in-expectation \textsc{Bounded-Square} mechanism by Lu and
Yu~\cite{LY08}.

\begin{algorithm}[ht]
\DontPrintSemicolon
\caption{\textsc{Exp-Bounded-Square}}\label{alg:exp-bounded-square}
\KwIn{Non-negative processing times $t = (t_{ij})_{i \in N, j \in M}$.}
\KwOut{Random allocation of tasks (set $M$) to machines (set $N$).}
\For{$j = 1, \dots, m$}{
	Draw $E_{ij} \sim \text{Exp}(1)$ independently for every $i \in N$. \;
	Draw an independent uniform random permutation $\pi_j$ of $N$. \;
	Set $t_j^{\min} := \min_{i \in N} \{ t_{ij} \}$ and $N_j := \{i \in N ~\colon~ t_{ij} \le 2 \cdot t_j^{\min} \}$. \;
	\eIf{$t_j^{\min} = 0$}{
		Assign $j$ to the first machine of $N_j$ in $\pi_j$.\;
	} {
		Assign $j$ to $\arg \min_{i \in N_j} \left\{ E_{ij} \cdot (t_{ij})^2 \right\}$. Break ties by $\pi_j$. \;
	}
}
\end{algorithm}

The mechanism processes tasks independently using a two-stage randomized assignment rule. 
First, it restricts the allocation to an eligible set of machines $N_j$ whose processing times are within a factor of $2$ of the minimum that any machine reports.
This pruning step prevents the mechanism from assigning a task to a highly inefficient machine. 
Then, if $t_j^{\min} = 0$, the task is allocated uniformly at random among those machines. Otherwise, it is assigned to an eligible machine that minimizes the \emph{exponentially scaled quadratic} processing time $E_{ij} \cdot (t_{ij})^2$, where $E_{ij} \sim \text{Exp}(1)$ are independent random weights drawn from the exponential distribution with rate parameter 1.
This quadratic scaling biases the allocation toward faster machines and was first used for the fractional case in~\cite{CKK10}. 
$\pi_j$ ensures consistent, input-independent tie-breaking.
For every fixed realization of the random weights and permutations, the allocation rule is deterministic and monotone, ensuring universal truthfulness with appropriate threshold payments (\Cref{sec:universal-truthfulness}). The exponential weights and the uniform permutations make the allocation probabilities of the mechanism match those of \textsc{Bounded-Square} (\Cref{lem:probabilities-exp-bounded-square}).

We first show in the next lemma that the assignment of each task follows
the exact same probability distribution as the fractional solution
computed by \textsc{Bounded-Square}~\cite{LY08}. Given any
instance $t$, for every task $j \in M$, let
$t_j^{\min} := \min_{i \in N} \{ t_{ij} \}$ denote the minimum
processing time for task $j$, and let $N_j := \{i \in N \;\colon\; t_{ij} \le 2 \cdot t_j^{\min} \}$
denote the set of \emph{eligible machines} for task $j$, as described in \Cref{alg:exp-bounded-square}.

\begin{lemma}
	\label{lem:probabilities-exp-bounded-square}
	Given any instance $t$, \textsc{Exp-Bounded-Square} allocates every task $j \in M$ independently to an eligible machine $i \in N_j$ with probability
	\[
		\begin{cases}
			\displaystyle\frac {1/(t_{ij})^2} {\sum_{s \in N_j} 1/(t_{sj})^2} &, ~\text{if}~ t_j^{\min} > 0 ~,\\
			1 / {|N_j|} &, ~\text{else}~,
		\end{cases}
	\]
	and to a non-eligible machine $i \notin N_j$ with probability 0.
\end{lemma}
\begin{proof}
	It follows directly from \Cref{alg:exp-bounded-square} that the allocation is task-independent, and every task $j \in M$ with $t_j^{\min} = 0$ is allocated to an eligible machine uniformly at random.
	Therefore, we consider any task $j \in M$ with $t_j^{\min} > 0$.
	Let $E_{ij} \sim \text{Exp}(1)$ (i.i.d.) and $X_i := E_{ij} \cdot (t_{ij})^2$ for every $i \in N_j$.
	Task $j$ is allocated to a machine from $\arg\min_{i \in N_j} \{X_i\}$ by the algorithm.
	It holds $X_i \sim \text{Exp}(1/(t_{ij})^2)$ independently for every $i \in N_j$ by definition of the exponential distribution, and thus 
	$\mathbb P(X_i = \min_{s \in N_j} \{X_s\}) = \frac {1/(t_{ij})^2} {\sum_{s \in N_j} 1/(t_{sj})^2}$.
\end{proof}

\subsection{Universal Truthfulness}
\label{sec:universal-truthfulness}

Next, we establish that the mechanism is indeed universally truthful.
We give a task-wise threshold payment scheme in \Cref{sec:payment-scheme-upperbound} for completeness.

\begin{lemma}
	\textsc{Exp-Bounded-Square} is a universally truthful mechanism.
\end{lemma}
\begin{proof}
	Fix the realizations of all $E_{ij}$ and $\pi_j$. 
	This yields a deterministic allocation rule, and we show that it is truthful for every such realization.
    The rule is task-independent, and each machine's processing cost is additive over tasks. 
    Hence it suffices to show that, for every task $j$, the allocation of $j$ is monotone in each machine's own value for $j$: then the task-wise threshold payments of \Cref{sec:payment-scheme-upperbound} make the mechanism truthful~\cite{AT01,Mye81}.
    
	Consider the allocation of a fixed task $j \in M$, and fix the reported processing times for $j$ for all machines except machine $i^* \in N$.
	Suppose $j$ is allocated to machine $i^*$ when reporting $t_{i^*j} > 0$.
	Then it holds $t_j^{\min} > 0$ and $i^* \in \arg \min_{i \in N_j} \left\{ E_{ij} \cdot (t_{ij})^2 \right\}$.
	Suppose now $i^*$ reports $t_{i^*j}' < t_{i^*j}$ instead. 
	The minimum reported processing time cannot be increased, and the new set of eligible machines must be a (weak) subset of the former $N_j$ that still includes machine $i^*$.
	If the minimum reported processing time is now 0, then clearly $t_{i^*j}' = 0$, and $i^*$ is the only eligible machine. Thus $j$ is still allocated to $i^*$.
	Otherwise, it holds $E_{i^*j} \cdot (t_{i^*j}')^2 \le E_{i^*j} \cdot (t_{i^*j})^2$, while all other scaled processing times are unchanged.
	Because $\pi_j$ is also fixed, the task $j$ is still allocated to machine $i^*$.
	Note that there is nothing to check when $t_{i^*j} = 0$: Since processing times are non-negative, no report $t_{i^*j}' < t_{i^*j}$ would exist.
\end{proof}

\subsection{Approximation}
\label{sec:approximation}

Here we show an upper bound on the expected makespan of \textsc{Exp-Bounded-Square}.
For any given instance $t$, let $R_i(t)$ denote the random set of tasks that is allocated to machine $i \in N$, and let $\theta_i(t) := \sum_{j \in R_i(t)} t_{ij}$ denote its total (random) processing time.
Moreover, let 
\[
	\Mech{\textsc{EBS}}{t} := \max_{i \in N} \theta_i(t)
\] 
denote the makespan of the random allocation obtained by \textsc{Exp-Bounded-Square} under $t$, and let
\[
	\mu(t) := \frac {n+1} 2 \cdot \Opt{t} \enspace,
\]
where $\Opt{t}$ is the optimal makespan under $t$.

We record two simple observations.
Every task $j \in M$ has a processing time of at least $t_j^{\min} = \min_{i \in N} \{ t_{ij} \}$ on every machine.
Hence $\max_{j \in M} t_j^{\min} \le \Opt{t}$.
Consequently, every allocation has a total processing time of at least $\sum_{j \in M} t_j^{\min}$, so there must be some machine with a load of at least $\frac 1 n \cdot \sum_{j \in M} t_j^{\min}$, which implies $\sum_{j \in M} t_j^{\min} \le n \cdot \Opt{t}$. 
We may also assume that there exists a task $j \in M$ with $t_j^{\min} > 0$.
Otherwise, \textsc{Exp-Bounded-Square} allocates every task $j \in M$ to a machine $i \in N$ that has processing time $t_{ij} = 0$ for it, and it holds $\Mech{\textsc{EBS}}{t} = 0$ with probability 1.
In this case, the allocation would be optimal.
As a consequence, it holds $\Opt{t} > 0$.
	
\begin{observation}
	\label{observation:opt-bounds}
	It holds $0 < \max_{j \in M} t_j^{\min} \le \Opt{t}$
	and $0 < \sum_{j \in M} t_j^{\min} \le n \cdot \Opt{t}$.
\end{observation}

Lu and Yu \cite{LY08} gave a bound on the maximum expected processing time of the machines, which also holds for our algorithm due to identical (fractional) probabilities of the allocation.
For completeness, we provide a formal proof in~\Cref{sec:expected-load-lu-yu-upperbound}.
Note that this result does not directly imply a bound for the {\em expected makespan} $\mathbb E[\Mech{\textsc{EBS}}{t}]$, which we aim to obtain.
\begin{lemma}[\cite{LY08}, Lemma 3]
	\label{lem:lu_yu_max_expected_load}
	For any instance $t$, it holds
	\[
		\mathbb E[\theta_i(t)] \le \mu(t) = \left(\frac{n+1}{2}\right)\cdot \Opt{t}
	\] 
	for every machine $i \in N$.
\end{lemma}

Using Hoeffding's inequality~\cite{Hoeffding1963}, we obtain the following lemma stating that the completion time of a machine deviates above its expected value with exponentially small probability. The formal proof is deferred to \Cref{sec:prob-deviation-completion-time}.
With a union bound, we can finally bound the expected makespan.
\begin{lemma}
	\label{lem:upper-bound-machine-i}
	Consider any instance $t$ with $\Opt{t} > 0$.
	For every machine $i \in N$ and for every $q \in \mathbb R_{\ge 0}$, it holds
	\[
		\mathbb P \left( \theta_i(t) \ge \mu(t) + q \cdot \Opt{t} \right) \le e^{ - q^2 / (2 n) } \enspace .
	\]
\end{lemma} 
\begin{theorem}
	\label{thm:upper-bound}
	For every instance $t$ and every $q \in \mathbb R_{\ge 0}$, \textsc{Exp-Bounded-Square} computes a random allocation with an expected makespan of
	\[
		\mathbb E[\Mech{\textsc{EBS}}{t}] \le \Opt{t} \cdot \left( \frac {n+1} 2 + q + 2 n^2 \cdot e^{ - {q^2} / {(2 n)} } \right) \enspace,
	\]
	where $\Opt{t}$ is the optimal makespan of instance $t$.
\end{theorem}
\begin{proof}
	By a union bound, \Cref{lem:upper-bound-machine-i} implies for every $q \in \mathbb R_{\ge 0}$ that
	\begin{equation}
		\label{ineq:union-bound-makespan}
		\mathbb P \left( \max_{i \in N} \theta_i(t) \ge \mu(t) + q \cdot \Opt{t} \right) \le n \cdot e^{ - q^2 / (2 n) } \enspace.
	\end{equation}
	By definition of \textsc{Exp-Bounded-Square}, it follows
	\[
		\theta_i(t) = \sum_{j \in R_i(t)} t_{ij} \le  \sum_{j \in R_i(t)} 2 t_j^{\min} \enspace.
	\]
	\Cref{observation:opt-bounds} yields a trivial upper bound on the random makespan of \textsc{Exp-Bounded-Square}:
	\[
		\Mech{\textsc{EBS}}{t} = \max_{i \in N} \theta_i(t) \le
		\sum_{j \in M} 2 t_j^{\min}
		\le 2 n \cdot \Opt{t} \enspace.
	\]
	We use this trivial upper bound for the event $\Mech{\textsc{EBS}}{t} \ge \mu(t) +  q \cdot \Opt{t}$.
	Recall that we bounded the probability of that event already in~\Cref{ineq:union-bound-makespan}.
	Therefore,
	\begin{align*}
		\mathbb E[\Mech{\textsc{EBS}}{t}] 
		\le &~ \mathbb P\left (\Mech{\textsc{EBS}}{t} < \mu(t) + q \cdot \Opt{t} \right) \cdot \left(\mu(t) + q \cdot \Opt{t} \right)\\
		& + \mathbb P\left (\Mech{\textsc{EBS}}{t} \ge \mu(t) + q \cdot \Opt{t} \right) \cdot 2n \cdot \Opt{t}\\
		\le &~ \mu(t) + q \cdot \Opt{t} + \mathbb P\left (\Mech{\textsc{EBS}}{t} \ge \mu(t) + q \cdot \Opt{t} \right) \cdot 2n \cdot \Opt{t}\\
		\le &~ \mu(t) + q \cdot \Opt{t} + n \cdot e^{ - q^2 / (2 n) } \cdot 2n \cdot \Opt{t}\\
		= &~  \Opt{t} \cdot \left(  \frac {\mu(t)} {\Opt{t}} + q + 2 n^2 \cdot e^{ - q^2 / (2 n) } \right) \enspace.
	\end{align*}
	Since $\mu(t) = \frac {n+1} 2 \cdot \Opt{t}$, the theorem follows.
\end{proof}

By rewriting $2n^2 \cdot e^{ - q^2 / (2 n) } = 2e^{2 \ln n - q^2 / (2n)}$, we finally obtain the claimed upper bound on the (expected) approximation ratio as a corollary, by selecting $q = 2 \sqrt{n \ln n}$.

\begin{corollary}\label{cor:upper-bound}
	For every instance $t$, it holds
	\[
		\mathbb E[\Mech{\textsc{EBS}}{t}] \le \Opt{t} \cdot \left( \frac {n+1} 2 + 2 \sqrt{n \ln n} + 2 \right) \enspace.
	\]
\end{corollary}

\appendix
\section{Proof of the padding lemma}
\label{app:padding}

We prove \cref{lem:padding}. Fix $\ell\geq2$. For every positive
integer $K$, we construct from $R$ a truthful mechanism on the core
tasks of $\mathcal{MC}_\ell$, the padded mechanism
$(x^{(K)},P^{(K)})$; the mechanism of the lemma is a limit of these as
$K\to\infty$.

\paragraph{Roadmap.}
\Cref{app:finite} constructs the padded mechanism and proves its
properties: truthfulness and symmetry~(iv)
(\cref{cl:truth,cl:sym}), and, up to errors that vanish as
$K\to\infty$, the load bound~\eqref{eq:padding-load}, eligibility~(i),
unique zeros~(ii) and zero-type irrelevance~(iii)
(\cref{cl:load,cl:elig,cl:zeros,cl:zti}). \Cref{app:limit,app:real} take the limit $K\to\infty$ at rational graph
inputs and extend it to all real graph inputs (\cref{cl:limit,cl:real}),
which proves the lemma. \Cref{app:witness} transfers a violation by the
limit mechanism to a padded instance of some $R_m$ (\cref{cl:witness}).

\subsection{The padded mechanism and its properties}
\label{app:finite}

For a fixed $K$, we pad every graph input of $\mathcal{MC}_\ell$ with
reservoir tasks, symmetrize the corresponding mechanisms $R_m$ within
classes, restrict them to the core tasks, and average over the
numbers of reservoir tasks, obtaining the padded mechanism $x^{(K)}$.
We then prove its properties in
\cref{cl:truth,cl:sym,cl:load,cl:elig,cl:zeros,cl:zti}.

\paragraph{Padding.}
Enumerate the positive rationals as $r_1,r_2,\ldots$, and let
\[
 T_K=\{(0,0)\}\cup\{(0,r_h),(r_h,0):1\leq h\leq K\}
\]
be a finite list of zero types, where a pair gives the values at the two
endpoints in a fixed order; every rational zero type lies in $T_K$ for
all large $K$. Let $d=\binom n2|T_K|$ and $B_K=\{K,\ldots,2K\}^d$, and
index the entries of a vector $\kappa\in B_K$ by a pair $\{i,j\}$ of
machines and a type $\tau\in T_K$. For a graph input $t$ of
$\mathcal{MC}_\ell$ and $\kappa\in B_K$, the padded instance $(t,\kappa)$
consists of
\begin{itemize}
\item the core tasks, with endpoint and owner values given by $t$;
\item for each pair $\{i,j\}$ and $\tau\in T_K$, $\kappa_{ij,\tau}$
 reservoir tasks between $i$ and $j$ with endpoint values $\tau$;
\item an extra self-loop at machine~$1$ with owner value $\beta_K=1/K$,
 which makes the optimum positive, so that~\eqref{eq:integral-bound}
 applies.
\end{itemize}
Every task has value $H_K=K^2$ at every machine other than its
endpoints or owner. Let $m_K(\kappa)$ be the number of tasks of
$(t,\kappa)$. We label the tasks of $(t,\kappa)$ so that the task set
and its partition into classes depend only on the number of reservoir
tasks in each class, not on their types.

\paragraph{Symmetrization.}
Fix a vector $\kappa\in B_K$, let $m=m_K(\kappa)$, and consider the
mechanism $R_m=(\hat x,\hat P)$. Let $\Pi_\kappa$ be the set of
permutations $\pi$ of the tasks of $(t,\kappa)$ that map every edge, core
or reservoir, to a parallel edge and fix every self-loop. For a
permutation $\pi$ of tasks and an input $s$, let
$(\pi s)_{ie}=s_{i,\pi(e)}$. The \emph{symmetrized mechanism}
$(\bar x,\bar P)$ for $\kappa$ is defined by the following allocation
function and payments, for every input $s$ with $m$ tasks:
\[
 \bar x_{ie}(s)=\frac1{|\Pi_\kappa|}\sum_{\pi\in\Pi_\kappa}\hat x_{i,\pi^{-1}(e)}(\pi s),
 \qquad
 \bar P_i(s)=\frac1{|\Pi_\kappa|}\sum_{\pi\in\Pi_\kappa}\hat P_i(\pi s).
\]

\begin{claim}[Symmetrized mechanism]\label{cl:sym-bar}
The symmetrized mechanism $(\bar x,\bar P)$ is truthful, and for every
$\gamma\in\Pi_\kappa$ and every input $s$,
\[
 \bar x_{ie}(\gamma s)=\bar x_{i,\gamma(e)}(s)
 \qquad\text{and}\qquad
 \bar P_i(\gamma s)=\bar P_i(s).
\]
\end{claim}

\begin{proof}
For a fixed $\pi$, the mechanism
$s\mapsto\bigl(\hat x_{i,\pi^{-1}(e)}(\pi s),\hat P_i(\pi s)\bigr)$ is
$R_m$ with the tasks renamed by $\pi$: machine $i$'s utility at $s$
equals its utility in $R_m$ at $\pi s$. It is therefore truthful, and so
is the average $(\bar x,\bar P)$, since the incentive
inequalities~\eqref{eq:ic} are linear in allocations and payments. For
the symmetry, substitute $\pi'=\gamma\circ\pi$, which ranges over
$\Pi_\kappa$ as $\pi$ does; then $\pi(\gamma s)=\pi's$ and
$\pi^{-1}(e)=\pi'^{-1}(\gamma(e))$, which gives both identities.
\end{proof}

\paragraph{The padded mechanism.}
Let $C$ be the set of core tasks, let $F_\kappa$ consist of the
remaining tasks of $(t,\kappa)$, and for a machine $i$ let $C_i\subseteq C$
be the core tasks of which $i$ is an endpoint or the owner. In
$(t,\kappa)$, machine $i$'s values on $C_i$ are given by $t$, and all its
other values depend only on $K$ and $\kappa$. For a graph input $t$ of
$\mathcal{MC}_\ell$, let
\[
 x^\kappa_{ie}(t)=\bar x_{ie}(t,\kappa)\quad(e\in C),
 \qquad
 P^\kappa_i(t)=\bar P_i(t,\kappa)-\sum_{f\notin C_i}(t,\kappa)_{if}\,\bar x_{if}(t,\kappa),
\]
where the sum ranges over all tasks of $(t,\kappa)$ outside $C_i$. This
is the slice of the mechanism $(\bar x,\bar P)$ to graph inputs of
$\mathcal{MC}_\ell$: as in \cref{sec:wmon}, each machine's payment is
reduced by its cost on its fixed values, so its utility is unchanged. The \emph{padded mechanism}
$(x^{(K)},P^{(K)})$ is the average of these restricted mechanisms over
all $\kappa\in B_K$:
\[
 x^{(K)}(t)=\frac1{|B_K|}\sum_{\kappa\in B_K}x^\kappa(t),
 \qquad
 P^{(K)}(t)=\frac1{|B_K|}\sum_{\kappa\in B_K}P^\kappa(t).
\]

\begin{claim}[Truthfulness]\label{cl:truth}
The padded mechanism $(x^{(K)},P^{(K)})$ is truthful.
\end{claim}

\begin{proof}
Each $(x^\kappa,P^\kappa)$ is a slice of the truthful mechanism
$(\bar x,\bar P)$ (\cref{cl:sym-bar}), hence truthful (\cref{sec:wmon}).
These mechanisms have the same domain, so their average is truthful by
the linearity of~\eqref{eq:ic}.
\end{proof}

\begin{claim}[Symmetry]\label{cl:sym}
Let $\gamma$ be a permutation of $C$ that maps each class $Q_{\{i,j\}}$
onto itself and fixes every self-loop. For every graph input $t$ of
$\mathcal{MC}_\ell$,
\[
 x^{(K)}_{ie}(\gamma t)=x^{(K)}_{i,\gamma(e)}(t)
 \qquad\text{and}\qquad
 P^{(K)}_i(\gamma t)=P^{(K)}_i(t).
\]
In particular, the padded mechanism satisfies the symmetry
property~(iv) of the lemma.
\end{claim}

\begin{proof}
Fix $\kappa\in B_K$. Extending $\gamma$ by the identity on $F_\kappa$
gives an element of $\Pi_\kappa$ with $\gamma(t,\kappa)=(\gamma t,\kappa)$.
By \cref{cl:sym-bar}, $x^\kappa_{ie}(\gamma t)=x^\kappa_{i,\gamma(e)}(t)$
for $e\in C$. The extension of $\gamma$ maps the tasks outside $C_i$ onto
themselves without changing machine $i$'s values on them, so by
\cref{cl:sym-bar} the cost subtracted in $P^\kappa_i$ is unchanged, and
$P^\kappa_i(\gamma t)=P^\kappa_i(t)$. Averaging over $\kappa$ gives both
identities. Machine $i$'s cost on $C_i$,
$\sum_{e\in C_i}(\gamma t)_{ie}\,x^{(K)}_{ie}(\gamma t)
=\sum_{e\in C_i}t_{i,\gamma(e)}\,x^{(K)}_{i,\gamma(e)}(t)$, is unchanged
too, and hence so is its disutility.
\end{proof}

\begin{claim}[Load bound]\label{cl:load}
Let $t$ be a graph input of $\mathcal{MC}_\ell$. For every
$\kappa\in B_K$, every machine's load under $(\bar x,\bar P)$ at
$(t,\kappa)$ is at most $\rho(C(t)+\beta_K)$. Consequently, for every
machine $i$,
\begin{equation}\label{eq:load-K}
 \sum_{e\in C}(t,\kappa)_{ie}\,x^{(K)}_{ie}(t)\leq\rho\,(C(t)+\beta_K),
\end{equation}
where $(t,\kappa)_{ie}$ does not depend on $\kappa$ for $e\in C$.
\end{claim}

\begin{proof}
Fix $\kappa\in B_K$. Allocating the core tasks by an eligible integral
allocation with makespan $C(t)$, each reservoir task to an endpoint with
value zero, and the extra self-loop to machine~$1$ shows
$\Opt{t,\kappa}\leq C(t)+\beta_K$; also $\Opt{t,\kappa}\geq\beta_K>0$,
since the extra self-loop has value at least $\beta_K$ at every machine.
The same holds for every relabelling $\pi(t,\kappa)$, $\pi\in\Pi_\kappa$.
Since we assume that $R_m$ has approximation ratio at most $\rho$
against the integral optimum at inputs with positive
optimum~\eqref{eq:integral-bound}, every machine's load under $R_m$ at
$\pi(t,\kappa)$ is at most
$\rho\,\Opt{\pi(t,\kappa)}\leq\rho(C(t)+\beta_K)$. Substituting
$f=\pi^{-1}(e)$ shows that machine $i$'s load under $(\bar x,\bar P)$ at
$(t,\kappa)$ is the average of these loads over $\pi\in\Pi_\kappa$,
which proves the first statement. Restricting the sum to $C$ and
averaging over $\kappa$ gives~\eqref{eq:load-K}.
\end{proof}

Since $\beta_K\to0$, \eqref{eq:load-K} gives the load
bound~\eqref{eq:padding-load} of the lemma as $K\to\infty$.

\begin{claim}[Eligibility]\label{cl:elig}
Let $t$ be a graph input of $\mathcal{MC}_\ell$, and let $e$ be a core
task and $i$ a machine that is neither an endpoint nor the owner of $e$.
Then
\[
 x^{(K)}_{ie}(t)\leq\frac{\rho\,(C(t)+\beta_K)}{K^2}.
\]
\end{claim}

\begin{proof}
Since $(t,\kappa)_{ie}=H_K=K^2$, this follows from~\eqref{eq:load-K}.
\end{proof}

As $K\to\infty$, \cref{cl:elig} gives eligibility~(i).

\begin{claim}[Unique zeros]\label{cl:zeros}
Let $t$ be a rational graph input of $\mathcal{MC}_\ell$, and let $e$ be
a core edge between $i$ and $j$ with $t_{ie}=0$ and $t_{je}=a>0$. If
$T_K$ contains the type of $e$, then
\[
 x^{(K)}_{je}(t)\leq\frac{\rho\,(C(t)+\beta_K)}{(K+1)\,a}.
\]
\end{claim}

\begin{proof}
Fix $\kappa\in B_K$. In $(t,\kappa)$, the edge $e$ has at least $K$
reservoir copies with identical columns, and swapping $e$ with one of
them is a permutation $\gamma\in\Pi_\kappa$ with
$\gamma(t,\kappa)=(t,\kappa)$. By \cref{cl:sym-bar}, $e$ and its copies
therefore receive the same fraction $z_\kappa=\bar x_{je}(t,\kappa)$ at
$j$, so machine $j$'s load at $(t,\kappa)$ is at least
$(K+1)\,a\,z_\kappa$. \Cref{cl:load} gives
$z_\kappa\leq\rho(C(t)+\beta_K)/((K+1)a)$, and averaging over $\kappa$
proves the claim, since $x^{(K)}_{je}(t)$ is the average of $z_\kappa$.
\end{proof}

Every rational zero type lies in $T_K$ for all large $K$, so as
$K\to\infty$, \cref{cl:zeros} gives unique zeros~(ii) at rational graph
inputs.

\begin{claim}[Zero-type irrelevance]\label{cl:zti}
Let $\tau\neq\sigma$ be rational zero types, and let $t$ and $t'$ be
rational graph inputs of $\mathcal{MC}_\ell$ that differ only in the
type of a core edge $e$ between $i$ and $j$: $\tau$ in $t$ and $\sigma$
in $t'$. If $T_K$ contains $\tau$ and $\sigma$, then for every core task
$f\neq e$ and every machine $u$,
\[
 \bigl|x^{(K)}_{uf}(t)-x^{(K)}_{uf}(t')\bigr|\leq\frac4{K+1}.
\]
\end{claim}

\begin{proof}
For $\kappa\in B_K$, let $\kappa'=\kappa+\one_{ij,\tau}-\one_{ij,\sigma}$.
If $\kappa'\in B_K$, the padded instances $(t,\kappa)$ and $(t',\kappa')$
have the same number of tasks in every class, so
$m_K(\kappa')=m_K(\kappa)$, and, by our labelling convention, the same
task set, the same $\Pi_\kappa$ and the same symmetrized mechanism. They also
contain the same multiset of task columns in every class:
$(t,\kappa)$ has $e$ of type $\tau$ and one
more reservoir task of type $\sigma$, while $(t',\kappa')$ has $e$ of
type $\sigma$ and one more of type $\tau$. Hence
$(t',\kappa')=\gamma(t,\kappa)$ for some $\gamma\in\Pi_\kappa$ that fixes
every core task other than $e$, and \cref{cl:sym-bar} gives
$x^{\kappa'}_{uf}(t')=x^\kappa_{uf}(t)$ exactly.

This pairing matches all terms of the average defining
$x^{(K)}_{uf}(t)$ except those with $\kappa_{ij,\tau}=2K$ or
$\kappa_{ij,\sigma}=K$, and all terms of the average defining
$x^{(K)}_{uf}(t')$ except those with $\kappa'_{ij,\tau}=K$ or
$\kappa'_{ij,\sigma}=2K$. Each of these four layers has weight $1/(K+1)$
in $B_K$, independently of the dimension $d$,
and the unmatched terms lie in $[0,1]$, which proves the claim.
\end{proof}

As $K\to\infty$, \cref{cl:zti} gives zero-type irrelevance~(iii) at
rational graph inputs.
\paragraph{From the padded mechanisms to the lemma.}
\Cref{cl:elig,cl:zeros,cl:zti} give properties~(i)--(iii) only
approximately, with errors that vanish as $K\to\infty$, and
\cref{cl:zeros,cl:zti} only at rational graph inputs, since they need
the types involved to lie in $T_K$. The remaining subsections remove
these restrictions. \Cref{app:limit} takes a limit of the padded
mechanisms as $K\to\infty$, which makes all properties exact at rational
graph inputs. \Cref{app:real} extends the limit mechanism to all real
graph inputs, as the lemma requires. Since the resulting mechanism is a
limit, \cref{app:witness} shows that if it violates a ratio $\rho'$,
then so does some $R_m$ at a finite input.

\subsection{The limit at rational inputs}
\label{app:limit}

For a graph input $t$ of $\mathcal{MC}_\ell$, let $t_i$ be machine $i$'s
vector of values on $C_i$, and let
$V^{(K)}_i(t)=V^{(K)}_i(t_i;t_{-i})$ be its disutility in the padded
mechanism (\cref{def:disutility}).

\begin{claim}[Limit at rational inputs]\label{cl:limit}
There is an infinite set $\mathcal K$ of positive integers such that the
limits
\[
 x_{ie}(t)=\lim_{K\in\mathcal K}x^{(K)}_{ie}(t)
 \qquad\text{and}\qquad
 V_i(t)=\lim_{K\in\mathcal K}V^{(K)}_i(t)
\]
exist for every rational graph input $t$ of $\mathcal{MC}_\ell$, every
machine $i$ and every core task $e$. With
$P_i(t)=t_i\cdot x_i(t)-V_i(t)$, the mechanism $(x,P)$ on the rational
graph inputs of $\mathcal{MC}_\ell$ is truthful and
satisfies~\eqref{eq:padding-load} and~(i)--(iv).
\end{claim}

\begin{proof}
By \cref{cl:truth} and \cref{sec:wmon}, $x^{(K)}_{ie}(t)\in[0,1]$ and
$0\leq V^{(K)}_i(t)\leq\sum_{e\in C_i}t_{ie}$. There are countably many
rational graph inputs, so a diagonal argument gives $\mathcal K$.

For every $K$, the incentive inequalities~\eqref{eq:ic}, the
normalization $V^{(K)}_i(0;t_{-i})=0$, feasibility, and the symmetry of
allocations and disutilities in \cref{cl:sym} are equalities or
non-strict inequalities among finitely many of these numbers, so they
hold for $x$ and $V$. This gives~(iv). By the normalization,
$P_i(0,t_{-i})=0$, so $V$ is the disutility function of $(x,P)$. With
the opposing values fixed, a machine with values $a$ that reports $b$
obtains utility $-V_i(b)+x_i(b)\cdot(b-a)\leq-V_i(a)$, by the first
inequality of~\eqref{eq:ic}; so $(x,P)$ is truthful.

\Cref{cl:elig} gives~(i). By~(i), machine $i$'s load at $t$ is
$\sum_{e\in C_i}t_{ie}\,x_{ie}(t)$, and
$\sum_{e\in C_i}t_{ie}\,x^{(K)}_{ie}(t)\leq\rho(C(t)+\beta_K)$
by~\eqref{eq:load-K}; letting $K\to\infty$
gives~\eqref{eq:padding-load}. Every rational zero type lies in $T_K$
for all large $K$. Hence \cref{cl:zeros} gives $x_{je}(t)=0$ for an
edge $e$ with $t_{ie}=0<t_{je}$, and then $x_{ie}(t)=1$ by~(i) and
feasibility, which is~(ii); and \cref{cl:zti} gives~(iii).
\end{proof}

\subsection{Real inputs}
\label{app:real}

For $h\geq1$ and $a\geq0$, let $R_h(a)=\lceil h!\,a\rceil/h!$, and apply
$R_h$ to every value of a graph input. Then $R_h(0)=0$, $R_h(a)>0$ for
$a>0$, $0\leq R_h(a)-a<1/h!$, and $R_h(a)=a$ for every rational $a$ and
all large $h$.

\begin{claim}[Real inputs]\label{cl:real}
The mechanism of \cref{cl:limit} extends to a truthful mechanism
$(x,P)$ on all nonnegative real graph inputs of $\mathcal{MC}_\ell$
that satisfies~\eqref{eq:padding-load} and~(i)--(iv). Moreover, there is
a subnet $(h_\alpha)$ of $(h)$ such that, for every graph input $t$,
$x(t)$ and $V(t)$ are the limits of $x(R_{h_\alpha}(t))$ and
$V(R_{h_\alpha}(t))$.
\end{claim}

\begin{proof}
For a graph input $t$, $R_h(t)$ is rational, $x_{ie}(R_h(t))\in[0,1]$
and $V_i(R_h(t))\in[0,\sum_{e\in C_i}(t_{ie}+1)]$ for all $h$. By
Tychonoff's theorem, these numbers, over all graph inputs $t$, converge
along a common subnet $(h_\alpha)$; let $x(t)$ and $V(t)$ be the limits,
and define $P$ as in \cref{cl:limit}. For rational $t$, $R_h(t)=t$ for
all large $h$, so this agrees with \cref{cl:limit}.

Each property passes to the limit along $(h_\alpha)$. \emph{Incentive
inequalities:} if $(a,t_{-i})$ and $(b,t_{-i})$ differ only in machine
$i$'s values, their roundings have the same opposing values,
so~\eqref{eq:ic} holds between them, and $R_h(b)-R_h(a)\to b-a$.
\emph{Normalization:} $R_h(0)=0$. \emph{Feasibility and~(i)} hold at
every $R_h(t)$. Truthfulness then follows as in \cref{cl:limit}.
\emph{(ii):} rounding keeps zero values zero and positive values
positive. \emph{(iii):} if $t$ and $t'$ differ only in the type of one
edge, and both types are zero types, then $R_h(t)$ and $R_h(t')$ differ
at most in the type of that edge, and both of its types there are
rational zero types. \emph{(iv):} $R_h(\gamma t)=\gamma R_h(t)$.
\emph{\eqref{eq:padding-load}:} by~(i), machine $i$'s load at $R_h(t)$
is $\sum_{e\in C_i}R_h(t)_{ie}\,x_{ie}(R_h(t))$, which converges to its
load at $t$, and $C(R_h(t))\to C(t)$ because $C$ is continuous.
\end{proof}

\Cref{cl:real} proves the lemma.

\subsection{Finite witnesses}
\label{app:witness}

\begin{claim}[Finite witnesses]\label{cl:witness}
Let $\rho'\geq0$, and let $(x,P)$ be the mechanism of \cref{cl:real}.
If $\Mech{x}{t}>\rho'\,C(t)$ for some graph input $t$ of
$\mathcal{MC}_\ell$, then there are $m\geq1$ and an input $s'$ with $m$
tasks such that $\Opt{s'}>0$ and $\Mech{R_m}{s'}>\rho'\,\Opt{s'}$.
\end{claim}

\begin{proof}
Fix a machine $i$ whose load at $t$ exceeds $\rho'\,C(t)$. By
\cref{cl:real} and~(i), its load at $R_{h_\alpha}(t)$ converges to its
load at $t$, and $C(R_{h_\alpha}(t))\to C(t)$; so its load exceeds
$\rho'\,C(s)$ at some rational graph input $s=R_{h_\alpha}(t)$. By
\cref{cl:limit} and~(i), this load is the limit of
$\sum_{e\in C_i}s_{ie}\,x^{(K)}_{ie}(s)$ as $K\to\infty$ in
$\mathcal K$, and $\beta_K\to0$, so for some $K\in\mathcal K$,
\[
 \sum_{e\in C_i}s_{ie}\,x^{(K)}_{ie}(s)>\rho'\,(C(s)+\beta_K).
\]
The left side is the average over $\kappa\in B_K$ of
$\sum_{e\in C_i}s_{ie}\,\bar x_{ie}(s,\kappa)$. Each of these is at most
machine $i$'s load under $(\bar x,\bar P)$ at $(s,\kappa)$, which by the
proof of \cref{cl:load} is the average over $\pi\in\Pi_\kappa$ of its
loads under $R_m$ at $\pi(s,\kappa)$, where $m=m_K(\kappa)$. Hence, for
some $\kappa$ and $\pi$, machine $i$'s load under $R_m$ at
$s'=\pi(s,\kappa)$ exceeds $\rho'(C(s)+\beta_K)$. Since
$0<\beta_K\leq\Opt{s'}\leq C(s)+\beta_K$ (proof of \cref{cl:load}) and
$\rho'\geq0$, we get $\Mech{R_m}{s'}>\rho'\,\Opt{s'}$.
\end{proof}

\section{Two-task separation without continuity}
\label{app:two-measurable}

We prove \cref{lem:two}\,(b). Fix a truthful two-task mechanism with
complementary allocations and zero-type irrelevance, and use
$V_i,V_j,V_{i,h},V_{j,h}$ ($h=1,2$) as in \cref{sec:two}. The axis identities~\eqref{eq:axes}
hold pointwise, since their proof used only zero-type irrelevance and the
area formula. The disutility functions are Lipschitz in their own values and, by the
argument of \cref{lem:opponent}, nondecreasing in their opposing values.
Joint Borel measurability in all four values is not guaranteed. We
construct Borel versions, built from countably many continuous
functions of the own values, that agree with the original disutilities at every own
value for almost every opposing value. These versions suffice for the
Laplace-transform computation.\footnote{A posteriori, the original disutilities
differ from Borel functions only on a set of the form
$\R_{\geq0}^2\times\mathcal N$ with $\mathcal N$ null, so they are
Lebesgue measurable, although they need not be Borel measurable.}

\paragraph{The version.}
For $s\in(0,\infty)^2$ let
\[
 \widehat V_i(t;s)=\sup\bigl\{V_i(t;a):\ a\in\Q_{\geq0}^2,\ a_1<s_1,\ a_2<s_2\bigr\}.
\]
This is a supremum of countably many functions, each continuous on the
own value $t$ and with a rational condition on $s$, so it is jointly
Borel. To check concavity, fix $s$ and choose rational vectors
$a_h<s$ increasing coordinatewise to $s$. Monotonicity in the opposing
values gives
\[
 \widehat V_i(t;s)=\lim_{h\to\infty}V_i(t;a_h).
\]
Thus $\widehat V_i(\cdot;s)$ is a pointwise increasing limit of concave
functions, and is concave. It is also $1$-Lipschitz in $t$, with the same
growth bound, and nondecreasing in $s$.

\paragraph{Agreement for almost every $s$.}
Fix a rational $t$ and write $g(s)=\widehat V_i(t;s)$. Since $V_i$ is
nondecreasing in $s$, for every $\eta>0$,
\[
 g(s)\leq V_i(t;s)\leq g(s+\eta\one).
\]
Along every diagonal line $r\mapsto s_0+r\one$, the function $g$ is
nondecreasing, so it is right-continuous at all but countably many
points. Integrating over the family of parallel diagonal lines, $g$ is
right-continuous along the diagonal at almost every $s$, and there,
letting $\eta\downarrow0$, $\widehat V_i(t;s)=V_i(t;s)$.
For each rational $t$, there is therefore a null set of opposing
values outside which the two disutilities agree at that $t$.
Take the union of these exceptional sets over all
$t\in\Q_{\geq0}^2$, and call it $\mathcal N_{\rm agree}$.
This union is still null because there are only countably many
rational $t$. For every $s\notin\mathcal N_{\rm agree}$, the two
disutilities agree at all rational own values simultaneously. Both are
Lipschitz in the own values, so agreement extends to all real
own values:
\[
 \widehat V_i(t;s)=V_i(t;s)
 \qquad\text{for every }s\in(0,\infty)^2\setminus\mathcal N_{\rm agree}
 \text{ and every }t\in\R_{\geq0}^2.
\]
The order of quantifiers is essential: the exceptional set
$\mathcal N_{\rm agree}$ is independent of $t$.
The same construction gives $\widehat V_j$, and one-dimensional versions
$\widehat V_{i,h},\widehat V_{j,h}$ of the one-task disutilities, which agree with the
originals outside countable sets of opposing values. Taking suprema on
both sides of~\eqref{eq:axes} shows that the axis identities hold exactly
for the versions.

\paragraph{The derivative identities for the versions.}
For almost every choice of opposing values, each version agrees
with its original disutility function at every own value. On each such
slice their own derivatives therefore agree wherever they exist,
and the area formula identifies these derivatives with the
allocations. This applies to both machines and to the one-task
disutilities. Since the versions are jointly measurable and Lipschitz in
the own values, Fubini's theorem shows that~\eqref{eq:complement}
holds almost everywhere for the versions.

Define $A$ and $B$ from the versions as in Step~2 of the proof of
part~(a). The axis identities and growth bounds still hold.
Steps~2--4 of that proof therefore apply and give $A=B=0$ almost
everywhere in the four value coordinates. These steps use
measurability and absolute continuity in the differentiated
coordinates, but do not require joint continuity.

\paragraph{One exceptional set for all own values.}
By Fubini's theorem, there is a null set
$\mathcal N_{\rm sep}\subset(0,\infty)^2$ such that, for every
$s\notin\mathcal N_{\rm sep}$, we have $A(t,s)=0$ for almost every $t$.
Fix any such $s$. The function $A(\cdot,s)$ is continuous on
$\R_{\geq0}^2$, since each disutility entering its definition is
Lipschitz in the own values. If it were nonzero at any $t$, it
would remain nonzero on a relative neighbourhood of positive
two-dimensional measure, a contradiction. Hence $A(t,s)=0$
for every $t\in\R_{\geq0}^2$, including boundary values.
Using the axis identities for the versions, we obtain
\[
 \widehat V_i(t_1,t_2;s)
 =\widehat V_i(t_1,0;s)+\widehat V_i(0,t_2;s)
 \qquad\text{for every }t\in\R_{\geq0}^2.
\]

Now set $\mathcal N=\mathcal N_{\rm sep}\cup\mathcal N_{\rm agree}$.
This is a null set of opposing values. For every $s\notin\mathcal N$,
the displayed identity holds for all $t$, and the version agrees
with the original disutility at $t$, $(t_1,0)$, and $(0,t_2)$.
Consequently,
\[
 V_i(t_1,t_2;s)=V_i(t_1,0;s)+V_i(0,t_2;s)
 \qquad\text{for every }s\in(0,\infty)^2\setminus\mathcal N
 \text{ and every }t\in\R_{\geq0}^2.
\]
Thus one exceptional set of opposing values works for all own
values in the fixed two-task mechanism. Exchanging the two machines
gives the corresponding assertion for $V_j$ and proves
\cref{lem:two}\,(b).

\section{Uniqueness of the Laplace transform}
\label{app:laplace}

Step~4 of the proof of \cref{lem:two} uses the uniqueness property of
the multivariate Laplace transform. We prove the instance we need, using
polynomial approximation.

\begin{lemma}\label{lem:laplace}
Let $F:\R_{\geq0}^4\to\R$ be measurable, with
$|F(r)|\leq\sum_{\nu=1}^4r_\nu$. If
\[
 \int_{\R_{\geq0}^4}F(r)\,e^{-\theta\cdot r}\,dr=0
 \qquad\text{for every }\theta\in(0,\infty)^4,
\]
then $F=0$ almost everywhere.
\end{lemma}

\begin{proof}
Make the change of variables
$z_\nu=e^{-r_\nu}$, for $\nu=1,\ldots,4$, and define
\[
 G(z)=F(-\log z_1,\ldots,-\log z_4),
 \qquad z\in(0,1)^4.
\]
The linear growth bound on $F$ bounds $|G(z)|$ by
$\sum_{\nu=1}^4(-\log z_\nu)$.
In particular, $G$ is square integrable on the cube, since
$\int_0^1(-\log z)^2\,dz=2$.

For any nonnegative integers $m_1,\ldots,m_4$, choose the four
transform rates to be $m_1+1,\ldots,m_4+1$. The change of variables
then gives
\[
 \int_{(0,1)^4}G(z)\prod_{\nu=1}^4z_\nu^{m_\nu}\,dz
 =
 \int_{\R_{\geq0}^4}F(r)
       e^{-\sum_{\nu=1}^4(m_\nu+1)r_\nu}\,dr
 =0.
\]
Thus the integral of $G$ times any monomial is zero, and by linearity
the same holds for every polynomial.

Polynomials are dense in $L^2([0,1]^4)$: continuous functions are
dense in this space, and polynomials approximate continuous functions
uniformly on the cube. Hence there are polynomials $p_q$ with
$\|G-p_q\|_{L^2}\to0$. Since $\int Gp_q=0$, the Cauchy--Schwarz
inequality gives
\[
 0\leq\int_{(0,1)^4}G(z)^2\,dz
 =\int_{(0,1)^4}G(z)\bigl(G(z)-p_q(z)\bigr)\,dz
 \leq \|G\|_{L^2}\,\|G-p_q\|_{L^2}\longrightarrow0.
\]
It follows that $G=0$ almost everywhere. Changing variables back
gives $F=0$ almost everywhere.
\end{proof}

\section{Omitted Details from \texorpdfstring{\Cref{sec:upperbound}}{the Upper Bound Section}}
\label{sec:appendix-upper-bound}

\subsection{Payment Scheme}
\label{sec:payment-scheme-upperbound}

Given any instance $t$ for \textsc{Exp-Bounded-Square}, we define payments $p_{ij}$ for each $i \in N$, $j \in M$.
The overall payment that machine $i$ receives is then given by $P_i(t) := \sum_{j \in M} p_{ij}$, which defines a possible payment scheme such that the mechanism is universally truthful.

Consider any fixed task $j \in M$, and suppose $j$ is allocated to machine $i^*$.
Set $p_{ij} = 0$ for all other machines $i \neq i^*$, fix their reported processing times, and define
\[
	t_{-i^*}^{\min} := \min \{\; t_{ij} \;\vert\; i \in N,\;i \neq i^* \} \enspace.
\]
If $t_{-i^*}^{\min} = 0$, we set $p_{i^*,j} = 0$.
Otherwise, we set
\[
	p_{i^*j} = \sup \left\{z \in [0, 2 t_{-i^*}^{\min}] \;\colon\; E_{i^*j} \cdot z^2 \le E_{ij} \cdot (t_{ij})^2 \text{ holds for all } i \neq i^* \text{ with } t_{ij} \le 2 \min \{z, t_{-i^*}^{\min} \} \right\}.
\]
Thus, $p_{i^*j}$ is the supremum of the processing time that $i^*$ could report for task $j$ such that it is still allocated to $i^*$.
Since, for every realization of the random choices, the allocation of each task is monotone in each machine's own value for it (see~\Cref{sec:universal-truthfulness}), these threshold payments make every realization truthful~\cite{AT01,Mye81}, so the mechanism is universally truthful.

\subsection{Bounding the Expected Load of Each Machine}
\label{sec:expected-load-lu-yu-upperbound}

\begin{proof}[Proof of~\Cref{lem:lu_yu_max_expected_load}]
	Consider a fixed optimal allocation for $t$, and let $J_i$ denote the set of tasks that is allocated to machine $i \in N$ in that optimal allocation.
	For every task $j\in M$, let $t^{\min}_j=\min_{i\in N} t_{ij}$, and let $N_j$ denote the set of eligible machines for task $j$.
	We assume w.l.o.g. that $t^{\min}_j> 0$ for every task $j$: Tasks with $t^{\min}_j = 0$ are allocated to a machine with processing time 0 for it, which does not increase $\theta_i(t)$, and removing such tasks does not increase $\Opt{t}$.
	For the optimum makespan it holds 
	\[
		\Opt{t} = \max_{i\in N}\sum_{j\in J_i} t_{ij}.
	\]
	Due to \Cref{lem:probabilities-exp-bounded-square}, the expected running time of an arbitrary machine $i$ in an allocation computed by \textsc{Exp-Bounded-Square} for instance $t$ can be expressed by
	\[
		\mathbb E[\theta_i(t)] 
		= \sum_{s \in N} \sum_{j\in J_s} x_{ij}t_{ij}
		= \sum_{j\in J_i} x_{ij}t_{ij} + \sum_{\substack{s\in N,\\ s\neq i}}\, \sum_{j\in J_s} x_{ij}t_{ij},
	\]
	where 
	\[
		x_{ij} = 
			\begin{cases}
				 \frac{1/(t_{ij})^2}{\sum_{s\in N_j}1/(t_{sj})^2} &,~\text{if}~i \in N_j,\\
				 0 &,~\text{otherwise}.
			\end{cases}
	\]
	For the first sum, it holds $\sum_{j\in J_i} x_{ij}t_{ij} \leq \sum_{j\in J_i} t_{ij} \leq \Opt{t}$.
	Next, we show that for every $s\neq i$ it holds that $\sum_{j\in J_s} x_{ij}t_{ij} \leq \Opt{t}/2,$ which will yield 
	\[
		\mathbb E[\theta_i(t)] \leq \left(\frac{n+1}{2}\right)\cdot \Opt{t}
	\]
	as required.
	
		\begin{claim}
			For every $s\neq i$ and every $j\in J_s$, it holds that $x_{ij}t_{ij}\leq \frac{t_{sj}}{2}.$
		\end{claim}
		\begin{proof} 
			If $i \notin N_j$ then $x_{ij}=0,$ and the claim holds. 
			Otherwise $i\in N_j,$ and so $t_{ij}\leq 2 t^{\min}_j$ holds. 
			Regarding $s \neq i$, we distinguish three cases:

			\begin{enumerate}[label=(\arabic*):]
				\item It holds $t_{sj}\leq 2t^{\min}_j,$ that is, $s\in N_j$.
				
					Then, taking \emph{all sums and products over machines} in $N_j$
					\begin{align*}
						x_{ij}t_{ij}
						&=\frac{1/(t_{ij})^2}{\sum_{k\in N_j}1/(t_{kj})^2}\cdot t_{ij}\\
						&=\frac{\prod_{ k\neq i}t_{kj}^2}{\sum_{l\in N_j}\prod_{k\neq l}t_{kj}^2}\cdot t_{ij}\\
						&=\frac{t_{ij}t_{sj}\prod_{ k\neq i,s}t_{kj}^2}{\sum_{l\in N_j}\prod_{k\neq l}t_{kj}^2}\cdot t_{sj}\\
						&=\frac{t_{ij}t_{sj}}{t_{ij}^2+t_{sj}^2+\sum_{l\neq i,s}t_{ij}^2t_{sj}^2/t_{lj}^2}\cdot t_{sj}\\
						&\leq \frac{t_{ij}t_{sj}}{t_{ij}^2+t_{sj}^2}\cdot t_{sj}\\
						&\leq \frac{1}{2}\cdot t_{sj}.
					\end{align*}
					
					\item $t_{sj}\geq 2t^{\min}_j,$ and $t_{ij}>t_j^{\min}.$ 
					
					Then there exists a machine $k\in N_j$ with $k\neq i,s,$ and $t_{kj}=t_j^{\min}\leq t_{sj}/2.$ 
					By replacing $t_{sj}$ by $t_{kj}$ in the argument above we obtain
					$x_{ij}t_{ij} \leq t_{kj} / 2\leq t_{sj}/4$.
					
					\item $t_{sj}\geq 2t^{\min}_j,$ and $t_{ij}=t_j^{\min}.$
					
					Then we obtain directly $x_{ij}t_{ij}\leq t_{ij}=t_j^{\min}\leq t_{sj}/2.$
			\end{enumerate}
		\end{proof} 
	Finally, by summing over all tasks in $J_s$ we obtain $\sum_{j\in J_s}x_{ij}t_{ij}\leq \frac{1}{2}\sum_{j\in J_s}t_{sj}\leq \frac{1}{2} \Opt{t}.$
\end{proof}

\subsection{Bounding the Probability for Deviation from the Expected Completion Time}
\label{sec:prob-deviation-completion-time}

\begin{proof}[Proof of~\Cref{lem:upper-bound-machine-i}]
	Consider a fixed machine $i \in N$ for a fixed instance $t$.
	Recall that $R_i(t)$ is the (random) set of tasks allocated to machine $i \in N$ by the algorithm.
	For every task $j \in M$, let 
	\[
		X_j := 
		\begin{cases} 
			t_{ij} 	&,~\text{if}~j \in R_i(t),\\
			0		&,~\text{else}.
		\end{cases}
	\]
	By definition of the algorithm, the random variables $X_1, \dots, X_m$ are independent, and it holds $0 \le X_j \le 2 \cdot t_j^{\min}$ for all $j \in M$, where $t_j^{\min}$ is the minimum processing time that any machine has for task $j$.
	Then it holds $\theta_i(t) = \sum_{j \in M} X_j$, and by assumption, $\Opt{t} > 0$. For every $q \in \mathbb R_{\ge 0}$, Hoeffding's inequality~\cite{Hoeffding1963} yields
	\begin{equation}
		\label{ineq:hoeffding}
		\mathbb P \Big( \theta_i(t) - \mathbb E \left[ \theta_i(t) \right] \ge q \cdot \Opt{t} \Big) 
		\le \exp \left( - \frac {2 (q \cdot \Opt{t})^2} {\sum_{j \in M} (2 t_j^{\min})^2} \right)
		=  \exp \left( - \frac {q^2 \cdot (\Opt{t})^2} {2 \cdot \sum_{j \in M} (t_j^{\min})^2} \right) \enspace.
	\end{equation}
	Clearly, for any $j \in M$, it holds $t_j^{\min} \le \max_{j' \in M} t_{j'}^{\min}$.
	Thus,
	\begin{equation}
		\label{ineq:t_j_min_square_sum}
		\sum_{j \in M} (t_j^{\min})^2 
		\le \left(\max_{j \in M} t_j^{\min}\right) \cdot \sum_{j \in M} t_j^{\min} 
		\le \Opt{t} \cdot n \cdot \Opt{t} \enspace,
	\end{equation}
	where we used~\Cref{observation:opt-bounds}.
	Since $\mathbb E[\theta_i(t)] \le \mu(t)$ holds for every machine $i \in N$ due to \Cref{lem:lu_yu_max_expected_load}, we obtain
	\begin{align*}
		\mathbb P \left( \theta_i(t) \ge \mu(t) + q \cdot \Opt{t} \right) 
		&\le 
		\mathbb P \Big( \theta_i(t) \ge \mathbb E[\theta_i(t)] + q \cdot \Opt{t} \Big)\\
		&= 
		\mathbb P \Big( \theta_i(t) - \mathbb E[\theta_i(t)] \ge q \cdot \Opt{t} \Big)\\
		&\overset{(\ref{ineq:hoeffding})}\le
		\exp \left( - \frac {q^2 \cdot (\Opt{t})^2} {2 \sum_{j \in M} (t_j^{\min})^2} \right)\\
		&\overset{(\ref{ineq:t_j_min_square_sum})}\le
		\exp \left( - \frac {q^2 \cdot (\Opt{t})^2} {2 n \cdot (\Opt{t})^2} \right)\\
		&= e^{ - q^2 / (2 n) } \enspace.
	\end{align*}
\end{proof}

\section*{AI methodology statement}
This paper supersedes our earlier preprint~\cite{ACKKS26}, 
which was written without any AI assistance. That preprint
proved a linear lower bound for universally truthful mechanisms whose
distributions have discrete support, by analysing deterministic integral
mechanisms via Yao's principle. The present results grew out of it
through extensive discussions with AI models, principally ChatGPT~6
(OpenAI). We first extended the argument of~\cite{ACKKS26} to
truthful-in-expectation mechanisms, and then to deterministic fractional
mechanisms, which gives the lower bound in its present form
(\cref{sec:lb}). The AI models contributed to developing and checking
the proof, to the exposition and source preparation, and to the figures.

The present proof keeps the overall architecture of~\cite{ACKKS26}: the
multi-clique instances and the hard star inside them; the counting
argument that selects the root through pair weights summing to one;
auxiliary tasks with a zero-cost machine, which the approximation
guarantee forces to be allocated trivially; and the need for a
continuity condition at the leaf values. The main new ingredients are
the averaging over task relabellings and reservoir counts, which yields
zero-type irrelevance (\cref{lem:padding}); two-task separation
(\cref{lem:two}), which replaces the characterization of deterministic
two-machine, two-task mechanisms; the exact identity for a fixed star
(\cref{lem:retention}), which replaces the Box Theorem; and root values
close to one together with a unit self-loop at the root. Together they
raise the bound of~\cite{ACKKS26}, a linear bound with a much smaller
constant, to $(n+1)/2$.

The upper-bound presentation was adapted from the authors' earlier
sources with AI assistance in checking details. Claude Opus~5.5
(Anthropic) assisted in merging the manuscript sources of this version,
in restructuring and editing its exposition, and in checking its
consistency. All listed authors are responsible for the content of the
paper.

\bibliographystyle{alpha}
\bibliography{main}

\end{document}